\documentclass[useAMS]{mn2e_modmargin}
\usepackage{amsmath}
\usepackage{url}
\usepackage{amsfonts}
\usepackage{amsbsy}
\usepackage{subfigure}
\usepackage{verbatim}
\usepackage{amssymb}
\usepackage{amsbsy}
\usepackage{graphicx}

\newcommand{\x}{\ensuremath{\mathbf{x}}}

\newcommand{\B}{\ensuremath{\mathbf{B}}}

\renewcommand{\u}{\ensuremath{\mathbf{u}}}
\renewcommand{\d}{\ensuremath{\partial}}

\newcommand{\ex}{\ensuremath{\mathbf{e}_{x}}}
\newcommand{\ey}{\ensuremath{\mathbf{e}_{y}}}
\newcommand{\ez}{\ensuremath{\mathbf{e}_{z}}}

\title[Vortices and the VSI]
 {Vortices and the saturation of the vertical shear instability in protoplanetary disks}
\author[H. N. Latter \& J. Papaloizou]{Henrik N. Latter\thanks{E-mail:
    hl278@cam.ac.uk} \& John Papaloizou \\
 DAMTP, University of Cambridge, CMS, Wilberforce Road,
Cambridge CB3 0WA, UK}

\begin{document}

\maketitle

\begin{abstract}
If sufficiently irradiated by its central star, a protoplanetary 
 disks falls
into an equilibrium state exhibiting vertical shear. 
This state may be subject to a hydrodynamical instability,
the `vertical shear instability' (VSI), 
whose breakdown into turbulence transports
a moderate amount of angular momentum while also facilitating
planet formation, possibly via 
the production of
small-scale vortices. 
In this paper, we show that VSI modes 
(a) exhibit arbitrary spatial profiles and (b) remain nonlinear
solutions to the incompressible local equations, no matter their amplitude.
The modes are themselves subject to parasitic Kelvin-Helmholtz
instability, though the disk rotation significantly impedes the parasites 
and permits the VSI to attain large amplitudes (fluid velocities
$\lesssim 10\%$ the sound speed). This `delay' in
saturation probably explains the
prominence of the VSI linear modes in global
simulations. More generally, the parasites may set the amplitude of VSI turbulence in
strongly irradiated disks. They are also important in breaking the
axisymmetry of the flow, via the unavoidable formation of vortices. 
The vortices, however, are not aligned with the orbital plane 
and thus express a pronounced $z$-dependence. 
We also briefly demonstrate that the vertical shear has little effect
on the magnetorotational instability, whereas magnetic fields easily quench
the VSI, a potential issue in the
ionised surface regions of the disk and also at larger radii. 
\end{abstract}


\section{Introduction}

Large swathes of protoplanetary (PP) disks are too poorly ionised
for the magnetorotational instability (MRI) to be active, at least as
classically understood (Turner et al.~2014). In particular, 
their cold dense interior regions (between roughly 1 and 10 AU) 
could be entirely laminar,
though non-ideal MHD may produce large-scale fields and winds (Lesur et al.~2014, Bai 2014,
Simon et al.~2015). In the last 10-15 years 
this state of affairs has renewed interest in
hydrodynamical routes to turbulence in these `dead zones', and
various weak instabilities have been uncovered (or rediscovered) that
might operate there (Fromang and Lesur 2017). 
It is unlikely that any of these instabilities is particular
widespread, nor the solution
to the question of angular momentum transport, but under certain
circumstances they could be important, especially for solid dynamics
and planet formation. 

We focus on the vertical shear instability (VSI),
a process that, as it names suggests, exploits any vertical
shear supported by the disk. Indeed, irradiated PP disks generally
manifest baroclinic equilibria that exhibit
gentle vertical variation in the rotation rate 
(Nelson et al.~2013, Barker and Latter 2015). The VSI is a
close cousin of the Goldreich-Schubert-Fricke instability (GSFI; Goldreich
and Schubert 1967, Fricke 1968), and is similarly centrifugal in
nature. When stabile stratification is included it is also
double-diffusive, with instability restricted to lengthscales upon
which thermal diffusion is strong (negating the stable entropy
gradient) but viscosity is weak (thus unable to obstruct the
unstable angular momentum gradient).

While the GSFI probably saturates at too low
a level to be important in stars (James and
Kahn 1970, 1971, see also Caleo et al.~2016), the VSI may have greater
traction in astrophysical disks, even if
the intensity of its saturated
state 
remains uncertain, and no doubt depends on how forcibly 
the vertical shear is maintained.
Unlike the GSFI, the VSI's nonlinear evolution has been pursued almost solely in global
simulations (Nelson et al.~2013, Stoll and Kley 2014, Richard et
al.~2016, Stoll et al.~2017). These show the flow degenerating into 
small-scale turbulence near or upon the disk's surfaces, followed
by the emergence of large-scale inertial waves with little vertical
structure. While the former activity is probably the work of the local modes
described by Boussinesq analyses (Urpin and Brandenburg
1997, Urpin 2003), the latter waves are global modes that are best captured
in vertically stratified models (Nelson et
al.~2013, Barker and Latter 2015, Lin and Youdin 2015, McNally and
Pessah 2015). 
Interesting features that arise in some
simulations are vortices, whose properties (aspect
ratios, lifetimes) depend on the background gradients (Richard et
al.~2016). Because of their potential impact on solid dynamics, it is
important to understand the nature of their formation
and its connection to general VSI saturation. 
This is the main topic of this paper.

We work within the convenient framework of Boussinesq hydrodynamics
because it is especially responsive to analytical techniques.
We first reveal important properties of the small-scale VSI modes,
such as: (a) they are nonlinear solutions to the governing
equations 
(and
hence can achieve large amplitudes naturally), and (b)
they can manifest
\emph{arbitrary} spatial profiles. This gives us confidence that our
Boussinesq analysis
adequately approximates some of the nonlinear waves and other features 
appearing in global simulations. We also apply the linear theory to
simple low-mass protoplanetary disk models, finding in particular that the VSI
grows at near its maximum rate even at 1 AU on scales of order
$10^{-2}H$ or less (where $H$ is the disk scale height). Larger-scale body
modes may be subdominant here, but small-scale VSI turbulence is still
viable. 

Because the nonlinear VSI modes exhibit significant shear, and vorticity
extrema, they are subject to Kelvin-Helmholtz parasitic instability.
As a consequence, small-scale vortices are a robust and unavoidable outcome of the VSI's
evolution. The advent of both axisymmetric and non-axisymmetric
parasites is delayed until the VSI achieves a relatively large
amplitude, characterised by a Rossby number greater than 1 (velocities between 1
and 10 \% the sound speed). Up to that point the axisymmetric parasites are stabilised
by the disk's radial angular momentum gradient, while the
non-axisymmetric parasites are sheared out before they get large. 
These two effects explain why the linear VSI modes feature so prominently
in global simulations, at least initially. 

More generally,
the parasites may set the level of VSI turbulence, but only if the vertical
shear is vigorously enforced by the stellar radiation field. 
If the VSI is especially efficient vis-a-vis the stellar
driving, the saturated state will be low amplitude and
controlled by the competition between the VSI and the driving,
not the parasites. One could also imagine intermediate scenarios when
all the physical processes are important.

We finally include a short section exploring the influence of MHD.
Because of the vertical shear, magnetic equilibria have to be
constructed carefully in order to satisfy Ferraro's law. We find that
sufficiently strong magnetic fields stabilise the VSI:
when ideal MHD holds and 
the average vertical plasma beta ($\beta$) is less than roughly
$(R/H)^2\sim 400$ then no VSI modes of any length can grow. 
But the critical $\beta$
is far higher for short-scale modes, and this poses a
problem for the VSI in the relatively well ionised 
upper layers of PP disks --- locations in which it is thought to be
most prevalent. It may also be an issue at larger radii throughout
the vertical column. 
In contrast, vertical shear has
no impact on the fastest growing magnetorotational modes. 

The paper plan is as follows. In Section 2 we revisit the 
VSI analysis in a Boussinesq local model, showing that the linear
modes are also nonlinear solutions and calculating both axisymmetric
and non-axisymmetric examples, growth rates, and stability criteria. 
A short subsection applies these results to simple low-mass disk models and
makes some comparison with previous work. Section 3 analyses the
parasitic instabilities that may attack the nonlinear VSI modes,
examining axisymmetric and non-axisymetric parasites separately and 
for different limiting amplitudes of the underlying VSI. 
Section 4 presents a relatively brief generalisation to MHD, in which
the MRI makes an appearance. We draw our conclusions in Section 5.

\section{The vertical shear instability}
\subsection{Equations of Incompressible hydrodynamics}

In this section we describe the VSI in the
framework
of incompressible hydrodynamics, a model that successfully represents
flows that are subsonic 
and which support only
small thermodynamic variations (as is the case for shear/centrifugal
instabilities like the VSI). Consequently, lengthscales are assumed to
be much less than the disk scaleheight, and global disk features are not
described properly. 
However, the disk's large-scale vertical shear can be incorporated
into this model in a straightforward and consistent way as shown by
Latter and Papaloizou (2017). At this stage vertical stratification
(i.e. buoyancy), 
radiative cooling, and viscous diffusion are neglected. 
This need not be a problem for the VSI, however, as there is often
a range of lengthscales between the vast gulf separating the viscous
scales ($\sim 10$ km) and the stratification scale ($\sim H$), within
which thermal diffusion (or cooling) dominates and `cancels out' the
latter stabilising effect. The
system on these scales neither feels viscosity, stratification, nor
thermal diffusion itself. In Appendix A we reproduce an analysis with the
omitted physics to show that our main results are unaltered. 

In addition, we adopt the shearing sheet formalism, whereupon a small
patch of the protoplanetary disk centred at $R=R_0$ and $Z=Z_0\neq 0$, and 
orbitting with frequency $\Omega_0=\Omega(R_0,\,Z_0)$
is described using a co-rotating Cartesian reference
frame. In it $x$, $y$, and $z$ represent the local radial, azimuthal,
and vertical directions, with their origin the centre of the box.

The equation of motion in our
incompressible, vertically shearing, local model is
\begin{equation} \label{eq1}
\d_t\u + \u\cdot\nabla\u = -\frac{1}{\rho}\nabla P -2\Omega\ez\times\u +
                 \Omega^2 (3x  +2z q ) \ex,
\end{equation}
where $\u$ is fluid velocity, $\rho$ is the constant density, $P$ is
the pressure, and 
$$q= -R_0\frac{\d\ln\Omega }{\d z} $$
evaluated at the point $R=R_0$, $Z=Z_0$ (see Latter and Papaloizou,
2017). Finally, the
fluid velocity satisfies the incompressibility requirement:
\begin{equation}\label{eq2}
\nabla\cdot\u =0. 
\end{equation}

The parameter $q$ quantifies the degree of vertical shear at
the shearing sheet location, and is generally small; from the thermal wind
equation comes the estimate $q\sim (H/R)$ (Barker and Latter 2015). 
Note that as $Z_0$ increases,
so usually will $q$.
The angular momentum gradient $\nabla (R^2\Omega)$ in the
shearing sheet is the vector
\begin{equation}\label{angel}
\nabla\ell = \tfrac{1}{2}R_0\Omega_0(\ex -2q\ez),
\end{equation}
which points predominantly in the radial direction.

One of the attractions of these equations is that they are controlled
by a single parameter $q$, and have no natural lengthscale: the outer
scale is taken to infinity, while the viscous scale is taken to zero. 
This, of course, makes them difficult to directly apply to real
systems --- but makes analytical progress much easier.

\subsection{General nonlinear perturbations}

The set \eqref{eq1} and \eqref{eq2} admits the steady equilibrium state of linear radial and vertical shear:
\begin{equation}\label{eqm1}
\u = \u_0 =-\frac{1}{2}\Omega_0(3x + 2 z q)\ey, \qquad P=P_0,
\end{equation}
where $P_0$ is a constant. 
This equilibrium is then perturbed by disturbances $\u_1$ and
$P_1$, which must obey
\begin{align}
&\d_t \u_1 +\u_0\cdot\nabla\u_1 + \u_1\cdot\nabla\u_1 =
  -\frac{1}{\rho}\nabla P_1 -2\Omega_0 \ez\times \u_1 \notag \\
& \hskip4cm  + \Omega_0
  \left(\frac{3}{2}u_{1x} + q u_{1z}\right)\ey,\label{perturb}
\end{align}
and $\nabla\cdot\u_1=0$. 

We next introduce the wavevector $\mathbf{k}(t)$, which is potentially a
function of time, and
construct the independent variable $\xi=
\mathbf{k}\cdot\mathbf{x}$. We
assume that our perturbations depend on $\xi$ and $t$ in the
following way:
\begin{equation}\label{p1}
 \u_1= \bar{\u}(t) f(\xi)\, \qquad  P_1= \bar{P}(t)\left(\int
f(\xi)\,d\xi\right), 
\end{equation}
where $\bar{\u}$ and $\bar{P}$ are functions to be determined, while $f$ is an arbitrary
 differentiable function. 

Using suffix notation, it is straightforward to show that the
incompressibility condition becomes
\begin{equation}
\mathbf{k}\cdot\bar{\u} = 0,
\end{equation}
and thus the fluid is constrained to move perpendicularly to the 
vector $\mathbf{k}$ (oscillations are transverse). But then
\begin{align*}
(\u_1\cdot\nabla\u_1)_j &= f\bar{u}_i \frac{\d f}{\d x_i}\,\bar{u}_j, \\
                      &=
                      f\frac{df}{d\xi}(k_i\bar{u}_i)\bar{u}_j = 0,                  
\end{align*}
and the nonlinear terms in the equation of motion vanish, a
significant simplification. It also means that our solutions are concurrently
linear and nonlinear solutions.

Turning next to the equation of motion, the pressure
gradient term simplifies in the following way
$$ \nabla P_1 = \bar{P}\nabla \left(\int f d\xi\right)= 
             \bar{P}\,f\,\nabla\xi = \bar{P}\,f\,\mathbf{k}, $$
and the remaining terms in the advective derivative become
$$ \d_t\u_1 + \u_0\cdot\nabla\u_1 = f\d_t\bar{\u}+\bar{\u}\frac{df}{d\xi}
\left[ x_i \d_t k_i-(\tfrac{3}{2} x + qz)\Omega_0 k_y 
\right].$$
Up to now the form of $\mathbf{k}$ has been left free. We next
select a form so that
the square bracketed expression above is zero. This yields ODEs for the
components of $\mathbf{k}$:
\begin{align*}
\frac{dk_x}{dt} =\frac{3}{2}\Omega_0 k_y, \quad 
\frac{dk_y}{dt} =0, \quad
\frac{dk_z}{dt} =q\Omega_0 k_y.
\end{align*}
Immediately we see that $k_y$ is a constant, while $k_x$ and $k_z$
increase linearly with time:
\begin{align}
k_x &= k_x^0 + \tfrac{3}{2}\Omega_0 k_y t, \\
k_z &= k_z^0 + q\Omega_0 k_y t,
\end{align}
where $k_x^0$ and $k_z^0$ are constants. Thus non-axisymmetric
disturbances possess wavevectors that are
sheared out by the differential rotation in $r$ and $z$. Ultimately 
$\mathbf{k}$ is perpendicular to surfaces 
of constant $\Omega$; i.e. $k_x/k_z \to
3/(2q)$ as $t\to \infty$ (Balbus et
al.~2009, Balbus and Schaan 2012). 
Our solutions may then be understood as (doubly) shearing waves
but with \emph{arbitrary} spatial profiles: they need not take the
form of sinusoids familiar from previous work (cf.\ Johnson
and Gammie 2005, Hawley and Balbus 2006, Balbus and Schaan 2012, 
Caleo and Balbus 2016, Caleo et al.~2016), and yet they still
remain nonlinear solutions to the governing equations. This property
has nothing to do with the vertical shear, of course, but issues
directly from the incompressible shearing box approximation. 

Now every term in Eq.~\eqref{perturb} is proportional to $f$. 
The spatial variable hence can be eliminated, leaving a vector ODE
in terms of time:
\begin{align} \label{nonaxe}
\frac{d\bar{\u}}{dt}= -\mathbf{k}\frac{\bar{P}}{\rho}-2\Omega_0\ez\times\bar{\u}
                     +\Omega_0\left(\frac{3}{2}\bar{u}_x +q \bar{u}_z\right)\ey.
\end{align}
By taking the scalar product of  (\ref{nonaxe}) with ${\bf k}$   and  making 
use of  the time derivative of  the incompressibility condition
we obtain an expression for $P/\rho$  in the form
$$ \frac{P}{\rho}=\frac{\Omega_0}{k^2}\left(k_y \bar{u}_x+2k_x \bar{u}_y 
 +2q k_y\bar{u}_z\right),$$
 with $k^2= k_x^2+k_y^2+k_z^2.$
Equation \eqref{nonaxe} can be reduced to a
single scalar ODE of third order. In the special case of $q=0$ this
reduces to second order and can be solved in terms of special
functions (see for example, Johnson and Gammie 2005). 

In order to obtain complete solutions in the non-axisymmetric
case the system must be solved numerically. However, 
we can obtain the perturbations' asymptotic behaviour as $t \rightarrow
\infty$ analytically which is sufficient to assess their
ultimate stability. We find that for large $t
\gtrsim 1/(q^2\Omega_0)$, 
$\bar{\u} \sim  
t^\alpha\text{exp}\left(\tfrac{4}{3}\text{i}q\Omega_0 t\right)$,
where 
$$\alpha= -\frac{1}{2} \pm \left(\frac{5k_z^0}{6k_y}-\frac{5q
    k_x^0}{9k_y}\right)\text{i}.$$
Thus all non-axisymmetric modes decay algebraically, while oscillating
on an intermediate timescale $\sim 1/(q\Omega_0)$.
Details of this calculation are provided in Appendix B.

\subsection{\label{axiVSI}The axisymmetric VSI}

The above scaling fails when the disturbances are axisymmetric, $k_y=0$.
In this case $\mathbf{k}$ is a constant and $\bar{\u}$ and
$\bar{P}$ can be decomposed into temporal Fourier modes, greatly easing the
analysis. 
We set them proportional
to $\text{exp}(st)$, where $s$ is a growth rate, and the perturbation
equations become algebraic:
\begin{align}\label{lin1}
& s\bar{u}_x = -\frac{1}{\rho}k_x \bar{P} + 2\Omega \bar{u}_y, 
& s\bar{u}_y=-\frac{1}{2}\Omega \bar{u}_x + q\Omega \bar{u}_z, \\
& s\bar{u}_z = -\frac{1}{\rho}k_z \bar{P}, 
& k_x\bar{u}_x + k_z\bar{u}_z = 0. \label{lin2}
\end{align}
Eliminating the dependent variables obtains the dispersion relation
\begin{equation} \label{growthrate}
s^2 = -\frac{k_z^2}{k^2}\Omega^2\left(1+ 2q \frac{k_x}{k_z} \right)
=-\frac{k_z}{k^2}\Omega^2\left[\mathbf{k}\cdot(\nabla \ell)^\perp\right],
\end{equation}
where $(\nabla \ell)^\perp= 2q\ex+\ez$, and is thus a vector
perpendicular
to the angular momentum gradient (cf.\ Eq.~\eqref{angel}).
This relation reproduces Eq.~(31) in Urpin and Brandenburg (1998)
and Eq.~(36) in Nelson et al.~(2013). 
Instability is assured whenever $q \neq 0$ for perturbations
with suitably oriented wavevectors:
\begin{equation}\label{hydrostab}
q\frac{k_x}{k_z} < -\frac{1}{2}.
\end{equation}
Marginal stability, $s=0$, 
occurs when $\mathbf{k}\to \ex$ or when
$\mathbf{k}$ is parallel to the angular momentum gradient
$\nabla\ell$ (Knobloch and Spruit 1982), with 
 growth limited to wavevector orientations lying
between $\ex$ and $\nabla\ell$.
 Because
$\nabla\ell$ is almost radial, the VSI is thus restricted to a narrow
arc of wavevector orientations, spanning an angle of only
$\approx 2q$ above
the radial axis.

There is no characteristic lengthscale in the governing
equations, and so the growth rate can only depend on wavevector orientation. 
Viscosity or vertical structure will
introduce a lengthscale dependence. See Appendix A and Section 2.4 for more
details. Because non-axisymmetric disturbances always 
tend to orient themselves along the gradient of $\Omega$,
i.e. $k_x/k_z=2/(3q)$, 
they becomes stable after some
point in their evolution, and hence ultimately decay, as shown
explicitly at the
end of the last subsection. 

The peak growth rate can be obtained by maximising $s$ over
$k_x/k_z$. The critical $k_x/k_z$ may be obtained from the 
quadratic
$$ q(k_x/k_z)^2 + (k_x/k_z) - q =0.$$
In the natural limit of small $q$, asymptotic solutions are
$k_x/k_z=q,\,-1/q$. Only the latter yields a positive growth
rate, and this corresponds to
\begin{equation}
s \approx \Omega |q|,
\end{equation}
in agreement with previous analyses (Urpin and Brandenburg 1998, Urpin
2003). Because $q\ll 1$ the growth rate is
considerably less than the orbital frequency, and fastest growth
occurs when $k_x/k_z \approx -1/q$, corresponding to disturbances that
are radially narrow
and vertically elongated (as seen in numerical simulations).

\subsection{Stratification, cooling, and viscosity}

We now briefly summarise the full problem in which buoyancy, viscous
diffusion, and cooling is included. In protoplanetary disks, the
photon mean free path varies substantially: from a tiny fraction of
$H$ at 1 AU, to greater than $H$ at 100 AU (e.g.\ Lin and Youdin 2015,
Lesur and Latter
2017). As a consequence, radiative cooling takes different forms,
depending on the age/mass of the disk, radial and vertical location,
and on the lengthscale of the perturbations. We mainly adopt the
diffusion approximation, but are aware this is invalid for sufficiently
short-scale perturbations at radii less than about 10 AU, and for
almost all perturbations at larger radii. The viscosity we employ is
treated as molecular, though might also represent diffusion by weak 
pre-existing small-scale turbulence, as long as that turbulence in no way impedes
the VSI. It could also be a model for unavoidable grid diffusion in
numerical simulations. 
 
Two new parameters now appear, (a) the Richardson
number $n^2=N^2/\Omega^2$, where $N$ is the vertical buoyancy frequency, and (b)
the Prandtl number $\text{Pr}= \kappa/\nu$ where $\kappa$ is thermal
diffusivity
and $\nu$ viscosity. 

\subsubsection{Main results}

In Appendix A we derive a number of results and summarise them here. First
we find that the analysis of Section 2.1 to 2.3 holds on a broad range of
radial lengthscales $\lambda$
\begin{equation}
\sqrt{\frac{1}{q}\frac{\nu}{\Omega}}\, \ll\, \lambda\,
 \ll \, \sqrt{\frac{q}{n^2}\frac{\kappa}{\Omega}}. 
\end{equation}
Perturbations within this range are sufficiently long so that
 viscosity is weak, but
sufficiently short so that the stabilising buoyancy force 
can be negated by thermal diffusion. Note that the vertical
lengthscale of the modes will be $1/q \sim R_0/H$ longer than $\lambda$. 
The existence of this range is
assured if 
\begin{equation}
\text{Pr} \lesssim \frac{q^2}{n^2} \sim
\left(\frac{H}{r}\right)^2\left(\frac{\Omega}{N}\right)^2.
\end{equation}
Equation \eqref{growthrate} provides a good estimate of 
the growth rate on this range.
An important thing to note is that, within the confines
of the Boussinesq model, the VSI lengthscales are set by the
diffusivities. This has to be, because the instability is double
diffusive in nature.

On shorter scales the diffusion approximation breaks down and 
we may replace it with Newtonian cooling. Typically, the transition
between these two cooling regimes occurs within the plateau of maximum
growth, described above. The wavenumber of the transition can be
estimated from either $k\sim 1/l_\text{ph}$, where $l_\text{ph}$
denotes the photon mean fee path, or simply $k \sim 1/\sqrt{\tau\kappa}$, where $\tau$ is the
Newtonian cooling timescale. At the transition point the cooling rate, which had
been increasing with $k^2$, hits the constant Newtonian ceiling. See Lin
and Youdin (2015) for a detailed treatment of this transition. 

As shown in Appendix A,
maximum growth can be maintained in this regime when the cooling is
sufficiently
fast
\begin{equation} \label{coolcond}
\Omega\tau \ll q\left(\frac{\Omega^2}{N^2}\right),
\end{equation}
where $\tau$ is the cooling time.
A similar expression is obtained by Lin and Youdin (2015). 
Even if the cooling is slow, growth can be sustained but at a lower
rate, with instability quenched only for exceedingly long cooling
times
\begin{equation}
\Omega\tau > \frac{q^2}{n^2}\text{Re},
\end{equation}
where Re=$H^2\Omega/\nu$ is the Reynolds number. 

\subsubsection{Application to a low-mass disc }

We now see how these estimates fare when applied to 
models of PP disk structure. Consider the
less massive nebula used in Lesur and Latter (2017), in which 
the disk's surface density is $\propto 140 R_\text{AU}^{-1}$ g
cm$^{-2}$ and the temperature is equal to 280$R_\text{AU}^{-1/2}$ K.
Here $R_\text{AU}$ is disk radius in AU. Protoplanetary disks are
moderately stratified, with $n\lesssim 1$, and we set $q\sim
H/R\sim 0.05$. Lesur and Latter (2017) also estimate the photon mean
free path, which rises from a little over $10^{-4}H$ at 1 AU, to $\sim
10^{-2}H$ at 10 AU, before exceeding $H$ at 100 AU. Fluctuations with
lengthscales below this cool
at the Newtonian rate: less than $10^{-3}\Omega^{-1}$ for radii
between 1
and 10 AU, but rising to $5\times 10^{-2}\Omega^{-1}$ at 100 AU. Table
I summarises some of this information. 
 
\begin{table}
\begin{tabular}{ ||l|l|l|l||}
 \hline
  Radius & 1 AU & 10 AU & 100 AU \\
 \hline
\hline
$\Sigma /(\text{g cm}^{-2})$ & 140 & 14 & 1.4 \\
\hline
$l_\text{ph} / H$ & $10^{-4}$ & $10^{-2}$ & $>1$ \\
\hline
$\tau\Omega$ & $10^{-4}$  & $\lesssim 10^{-3}$ & $ 10^{-2}$ \\
\hline
$\lambda^\text{plat}_\text{VSI} / H$ & $3\times 10^{-2}$ & $\lesssim 1$ & $\sim
1$ \\
 \hline
\end{tabular}
\caption{Properties of a low-mass PP disk at three different
  radii. Here $\Sigma$ is surface density, $l_\text{ph}$ is photon
  mean path, and $\tau$ is the cooling rate for lengthscales less than 
  $l_\text{ph}$. Here $\lambda^\text{plat}_\text{VSI}$ denotes the
  lengthscale below which the VSI achieves its maximum value (in the
  diffusion approximation). On lengthscales less than $l_\text{ph}$, however,
  the VSI can grow at its maximum level if $\tau\Omega \ll 5\times
  10^{-2}$. }
\end{table}

The plateau
of maximum growth occurs for wavelengths roughly less than
$q^{1/2}n^{-1}\text{Pe}^{-1/2}H$, where Pe$=H^2\Omega/\kappa$ is the Peclet number.
At 1 AU we find that the plateau begins at a wavelength less than
$3\times 10^{-2}H$, and thus the fastest growing modes are shortscale
and well described by our Boussinesq model. The photon mean free path, 
on the other hand, is $l_\text{ph}\sim 10^{-4}H$ and so the diffusion
approximation holds for at least 2 decades in wavenumber upon which
maximum growth is achieved. Below  
$l_\text{ph}$, however, the VSI will remain growing at the same maximum
value until the viscous cut-off, this is because the Newtonian cooling
rate $\tau$ easily satisfies \eqref{coolcond}. 

At 10 AU, we find that the plateau
begins at a longer lengthscale $\lesssim H$, and thus the local
model is a poor approximation on the long end of the range of
maximum growth. At this location $l_\text{ph}\sim 10^{-2}$ and the diffusion
approximation still holds on this upper range, while the
Newtonian cooling rate still satisfies \eqref{coolcond}. 

At 100 AU,
$l_\text{ph}> H$ and the diffusion approximation must be discarded entirely 
in favour of a Newtonian cooling prescription (Lin and Youdin
2015). Moreover, maximum growth is no longer possible 
because $\Omega\tau \sim 10^{-2} \sim q(\Omega^2/N^2)$, and
\eqref{coolcond} no longer satisfied. At these radii the VSI begins to
suffer, growing at a lower rate, in rough agreement with analogous
calculations (Lin and Youdin 2015, Malygin et al.~2017).

To inspect how the VSI fares in more massive massive disc models the
reader is directed to Lin and Youdin (2015) and Malygin et al.~(2017).
Due to the longer cooling times of these objects, the VSI is
less prevalent. However, we do point out that short-scale VSI modes should
not be neglected in these analyses. Though they may be inefficient at
transporting angular momentum, or erasing the vertical shear, 
they are still perfectly good at generating
vortices. And vortices need not be large-scale to do interesting
things with solid particles.

\subsection{Example solutions}

\begin{figure}
\center
\includegraphics[width=8cm]{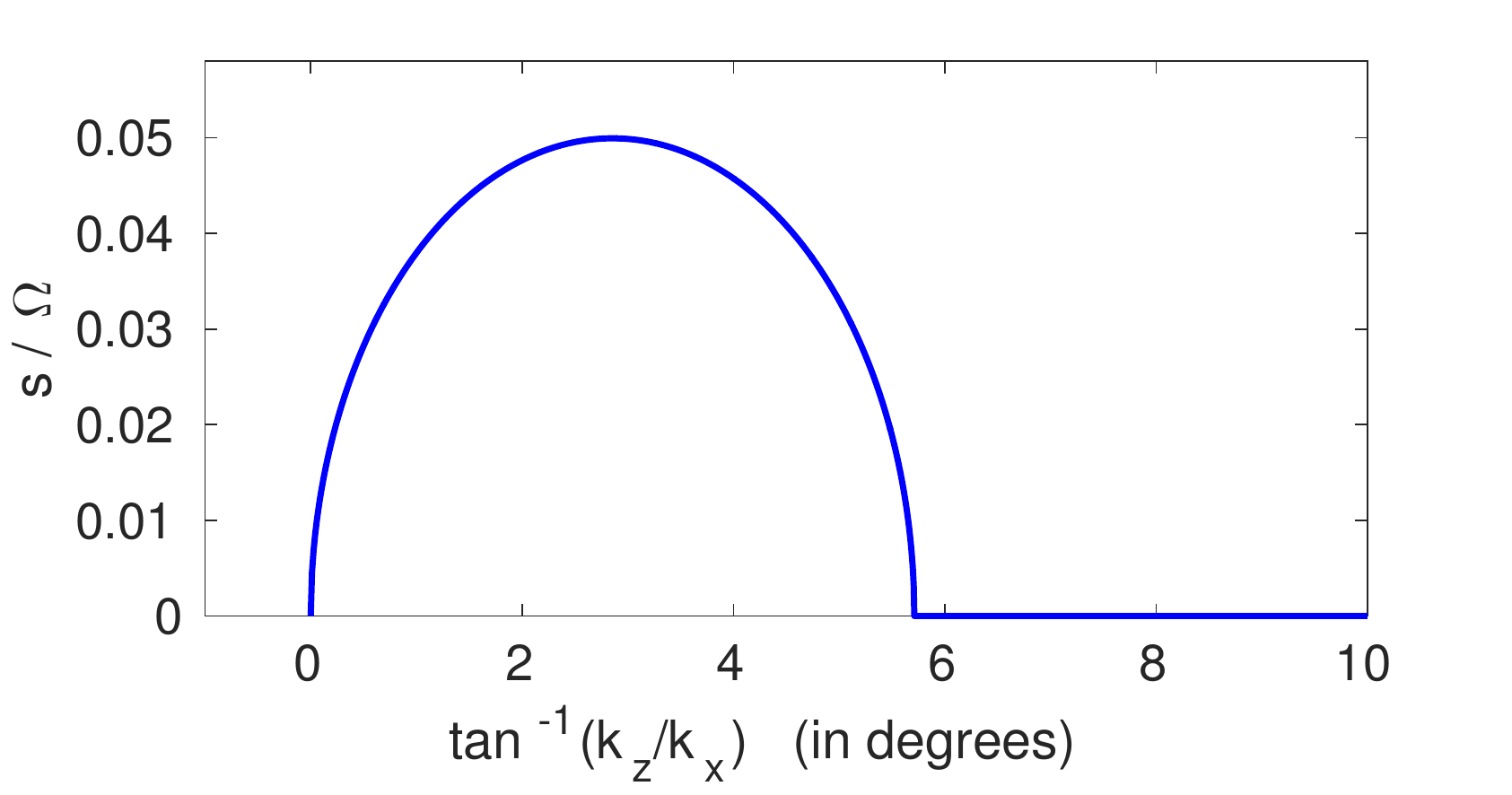}
\caption{Growth rate of the VSI, as determined by
  Eq.~\eqref{growthrate} as a function of wavenumber orientation angle
$\tan^{-1}(k_z/k_x)$ as measured from the $x$-axis. The angle is given
in degrees. Here $q=-0.05$ and we see that instability takes its
maximum value $\sim |q|\Omega$, while being restricted to a narrow arc
of wavevector angles, spanning only some $6^\circ$.}
\label{fig::VSI}
\end{figure}

In this subsection we present some numerical solutions describing the
VSI's evolution. To get a
better idea of how the unstratified axisymmetric mode depends on the
orientation of its wavevector, we plot the growth rate $s$ versus
$\tan^{-1}(k_z/k_x)$, which is the angle the wavevector makes with
respect to the $x$-axis. Quite clear is the extremely narrow range of
angles available to the VSI, some $6$ degrees above the disk plane.
Maximum growth, as shown earlier, occurs when $k_x/k_z\approx -1/q$. 

\begin{figure}
\center
\includegraphics[width=8cm]{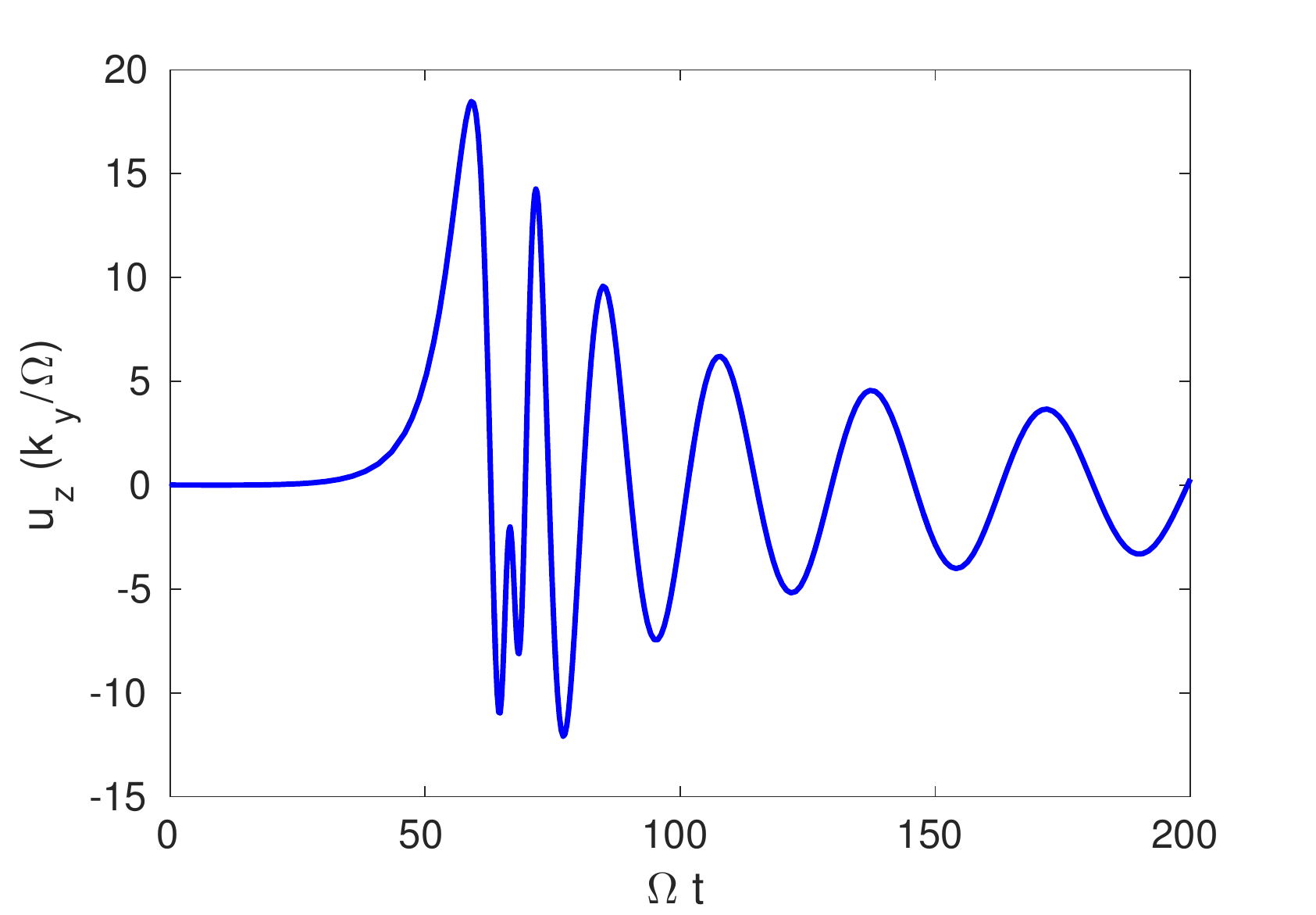}
\caption{Numerical solution to Eq.~\eqref{nonaxe} for a non-axisymmetric
disturbance with $k_{x0}/k_y=-100$, $k_{z0}/k_y=1$, and initial condition
$\bar{u}=(0.01,\,-0.01,\,0.01)$. Finally, $q=-0.1$. }
\label{fig::nonaxVSI}
\end{figure}

In Figure 2 we plot the evolution of $\bar{u}_x(t)$ for a non-axisymmetric VSI
disturbance in time. The system of equations \eqref{nonaxe} were
solved with a Runge-Kutta algorithm, for fiducial initial conditions
corresponding to a leading wave at $t=0$.
Substantial growth occurs near $\Omega t\approx 42.5$, as the wave
vector passes through the arc of axisymmetric growth (around
$k_x/k_z\approx -1/q$). Once the wavevector enters
stable orientations, at later times, the disturbance decays as
predicted,
oscillating with a frequency $\sim q\Omega$ with an envelope going as
$t^{1/2}$. 

\begin{figure}
\center
\includegraphics[width=8cm]{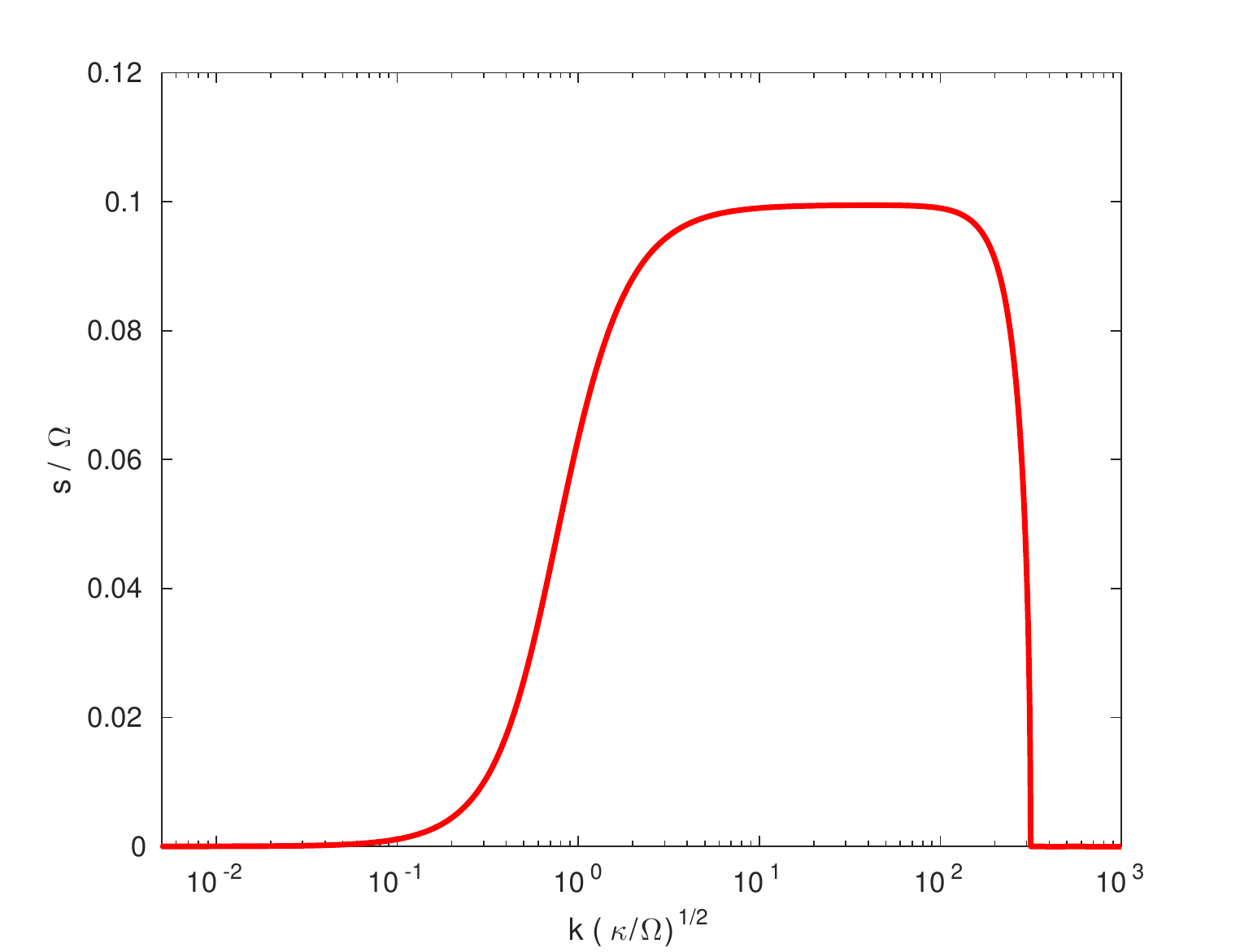}
\caption{Growth rate of the VSI when stratification, viscosity, and
  thermal diffusion are included. Parameters are $q=-0.1$,
  $\text{Pr}=10^{-6}$, $n^2=0.1$, and $k_x/k_z=10$. Here the viscous scale
  corresponds to $k_\nu\sim 10^3$, in units of $\sqrt{\Omega/\kappa}$,
  while the thermal diffusion scale is $\sim 1$. The fastest
growth occurs on a range of lengths between these limits, and here $s$
is approximately the same as if there were no stratification, nor
viscous and thermal diffusion.}
\label{fig::fullVSI}
\end{figure}

Finally, in Figure 3 we plot a representative
dispersion relation for the axisymmetric VSI with stratification,
viscosity, and thermal diffusion.
Note the conspicuous plateau extending from roughly the thermal
diffusion scale $\sqrt{\kappa/\Omega}$ to the viscous cut-off some two
orders of magnitude smaller. Upon the plateau $s$
takes almost the same value as in the inviscid unstratified situation,
Eq.~\eqref{growthrate}.

\section{Kelvin-Helmholtz parasites}

Because the VSI modes are nonlinear solutions
they can grow to arbitrarily large amplitudes, at least in principle. 
Before they grow too large,
however, they will be attacked by secondary parasitic
instability, especially if the spatial profile of the modes $f(\xi)$
exhibits sufficient shear. As in similar problems (e.g.\ the MRI), the
primary parasitic instability will be
of Kelvin-Helmholtz type (Goodman and Xu 1994, Latter et al.~2009,
Pessah and Goodman 2009), 
which in the hydrodynamic context may lead to vortex
formation. (In MHD, magnetic tension prevents the rolling up of vortex
sheets.) In this section of the paper we explore how effective
parasitic modes are and how associated vortices
may arise. 

We take the state of homogeneous vertical and
radial shear, $\u_0$ and $P_0$, given by Eq.~\eqref{eqm1} and then add to
it an \emph{axisymmetric} VSI mode,
$\u_1$ and $P_1$, given by Eq.~\eqref{p1}. The VSI velocity field may be
written as
\begin{align}
\u_1= V\,\bar{\u}\,f(\xi)\,\text{e}^{st},
\end{align}
where $\bar{\u}$ is a constant eigenvector determined from Eqs
\eqref{lin1}-\eqref{lin2}, nondimensionalised and 
normalised so that $|\bar{\u}|=1$, and $s$
is the growth rate, given by \eqref{growthrate}. Taking $f(\xi)$
to be dimensionless and with a maximum value of order unity, the
parameter $V$ is then the characteristic 
velocity amplitude of the VSI mode
at $t=0$. It possesses units of speed. Because the solution is
axisymmetric, the wavevector
$\mathbf{k}$ is a constant, which introduces the only lengthscale in
the problem. 

The key parameter in what follows is the
amplitude of the VSI mode, $V$, which can only be expressed in terms of 
$\Omega/k$. We regard the VSI mode to be of small amplitude if
$V\ll \Omega/k$, of moderate amplitude if $V\sim
\Omega/k$, and of large amplitude if $V\gg \Omega/k$. This motivates
the introduction of the VSI's Rossby number Ro$=kV/\Omega$, a proxy
for the VSI amplitude.

The sum of the two fields, $\u_0$ and $\u_1$,
we take to be our background, time varying, state. We next perturb
this state with small disturbances, denoted by $\u_2$ and $P_2$. 
The linearised equations for the disturbances are
\begin{align}
&\d_t\u_2 -(\tfrac{3}{2} x+ qz)\Omega\, \d_y \u_2 +\u_1\cdot\nabla\u_2
+(\mathbf{k}\cdot\u_2)\d_\xi\u_1 \notag \\
& \hskip0.5cm  =-\frac{1}{\rho}\nabla P_2 -2\Omega
\ez\times \u_2  + \left(\tfrac{3}{2}\Omega
  u_{2x}+q\Omega u_{2z}\right)\ey \label{p2}
\end{align}
and
\begin{equation}
\nabla\cdot\u_2=0.
\end{equation}
These differ from the usual linearised equations for a homogeneous
background
of uniform shear because of the additional terms involving $\u_1$.
The equation of motion \eqref{p2} depends explicitly on
time (through $\u_1$) and $x$ and $z$, thus we cannot Fourier decompose the
perturbation $\u_2$ in any of these independent variables. As such
Eq.~\eqref{p2} presents a considerable challenge to solve: it is a PDE in three
variables. In the next two subsections we look at various limits in which
the problem reduces to something more manageable. Subsequently, we
collate these results and construct a theory for the destruction of
VSI modes and the emergence of vortices from their breakdown.

\subsection{Axisymmetric parasites}

The first simplifying case we investigate assumes that the parasite is
axisymmetric, so that
$\d_y\u_2=\d_yP_2=0$. This is not especially restrictive: 
for modest amplitudes of the background VSI (and parasitic
growth rates of order $\Omega$ or less),
axisymmetric parasites should be the most effective in overcoming their
hosts. This is because non-axisymmetric disturbances are sheared
out on their instability timescale and only grow transiently: 
in their short window of growth they may fail to achieve sufficient amplitudes
(Appendix B of Latter et al.~2010, and also Rembiasz
et al.~2016).

When the parasites are axisymmetric, the spatial variables in \eqref{p1} appear
only via $f$'s dependence on $\xi=\mathbf{k}\cdot\x=k_x x + k_z z.$
To exploit this fact, we
 introduce the coordinate
transformation $(x,\,z)\to (\zeta,\,\eta)$, where the $\zeta$ coordinate
axis points in the direction of $\mathbf{k}$ and the $\eta$ coordinate
direction is perpendicular to $\mathbf{k}$. The transformation corresponds to a
rotation of the $(x,\,z)$ axis by an angle $\tan^{-1}(k_z/k_x)$.
Thus
\begin{align}
\zeta =\frac{x k_x+zk_z}{k}, \qquad \eta= \frac{zk_x-xk_z}{k}.\label{Newcoords}
\end{align}
Note that the earlier (dimensionless) variable $\xi$ is related to
$\zeta$ by the rescaling $\xi= k\zeta$. Thus $f(\xi)$ is a
straightforward
function of $\zeta$ alone.

On account of the coordinate change, the governing equations
lose their explicit $\eta$ dependence and we may express the
perturbations as
$$\u_2,\,P_2 \propto
\text{exp}(\text{i}K_\eta \eta ),$$
 where $K_\eta$ is a real
wavenumber
and for now the amplitudes  $\u_2,\,P_2 $ depend on $\zeta$ and time, $t.$
In components the evolution equations for the parasite are
\begin{align}\label{axieq1}
\text{i}\overline{\sigma} 
u_{2\zeta} &= -\d_\zeta h_2 + \frac{2\Omega k_x}{k} u_{2y},\\
\text{i}\overline{\sigma} 
 u_{2y} &= - (V\bar{u}_{y}\d_\zeta f)\,u_{2\zeta} -
\tfrac{1}{2}\Omega u_{2x} + q\Omega u_{2z}, \label{yeq}\\
\text{i}\overline{\sigma}
  u_{2\eta} &= - (V\bar{u}_{\eta}\d_\zeta
f)\,u_{2\zeta}-\text{i}K_\eta h_2 -\frac{2\Omega k_z}{k} u_{2y}, \label{etaeq}\\
&0= \d_\zeta u_{2\zeta} + \text{i}K_\eta u_{2\eta},\label{axieq2}
\end{align}
where $h_2= P_2/\rho_0$, and the operator
$$\overline{\sigma}= -\text{i}\,\d_t+ K_\eta\,V \bar{u}_{\eta} f(\xi).$$
Using simple trigonometry, 
the $x$ and $z$ components of $\u_2$ can be rewritten as
$$ u_{2x}=\frac{k_x u_{2\zeta} - k_z u_{2\eta}}{k}, \quad u_{2z}=\frac{k_z u_{2\zeta}+k_x
u_{2\eta}}{k}.$$

The equations \eqref{axieq1}-\eqref{axieq2} depend explicitly on $\zeta$ and $t$, the latter
through ${\bf u}_1$'s proportionality to $\exp(st)$. Its solutions,
as a consequence,
are not separable. However, in the special case of the VSI's marginal
stability $s=0$ (corresponding to $k_z=-2qk_x$),
 the system's time-dependence drops out and we
may let the parasitic perturbations be $\propto \exp({\rm i}\sigma
t),$ where $\sigma$ is a possibly complex frequency.
Consequently,
\begin{align}
\overline{\sigma} \rightarrow \sigma+ K_\eta\,V \bar{u}_{\eta} f(\xi),\label{separable}
\end{align}
and Eq.~(\ref{yeq}) transforms into the simpler
\begin{align}
&\hspace{-2mm} \text{i}\overline{\sigma} u_{2y} 
= - (V\bar{u}_{1y}\d_\zeta f)\,u_{2\zeta} -
\frac{1}{2}\Omega\sqrt{1+4q^2} u_{2\zeta} \equiv -\Lambda u_{2\zeta}
 \label{yeq1}
\end{align}

Strictly speaking, we are permitted to make the above assumptions and
manipulations
 only when $s=0$. But if $q$ is assumed to be small, $s\sim |q|$,
and we are concerned with a time interval much less than the VSI's
e-folding time, then we may neglect variations in the quantity
$\exp(st),$ in exactly the same way as if the VSI was marginal. Then
(\ref{separable}) and (\ref{yeq1}) hold, and
terms of order $q$ and higher may then be neglected in
  (\ref{axieq1})~-~(\ref{axieq2}). 

Via these manipulations and approximations we 
finally obtain a second-order eigenvalue problem in a single variable, $\zeta$, with
eigenvalue $\sigma$. The input parameters are $q$
(from which we obtain the ratio of the  components of $\u_1$), 
$V$, and the wavenumber of
the parasite $K_\eta$. In addition, the spatial structure of the host VSI
mode, $f(\xi)$, must also be supplied.

By employing (\ref{yeq}) to eliminate  $u_{2y},$ (\ref{axieq2}) to eliminate $u_{2\eta}$ 
and  then (\ref{etaeq}) to eliminate $h_2,$
the set \eqref{axieq1}-\eqref{axieq2}  reduces to a single
ODE, for $u_{2\zeta}$, which can be written in the form
\begin{align}
&\frac{d^2 u_{2\zeta}}{d\zeta^2} +\frac{4{\rm i}qK_\eta\Omega}{\sqrt{1+4q^2}\overline{\sigma}} 
\frac{d}{d\zeta}\left(\frac{\Lambda u_{2\zeta}}{\overline{\sigma}}\right)\nonumber\\
&-\left(K_\eta^2 + \frac{V K_\eta \bar{u}_{1\eta}\,(d^2 f/d\zeta^2)}{\overline{\sigma}}
-\frac{2\Omega K^2_\eta\Lambda}
{\overline{\sigma}^2\sqrt{1+4q^2}} 
 \right)u_{2\zeta}
=0,\label{aaxistabgov}
\end{align} 
Equation (\ref{aaxistabgov}) governs the stability of the VSI mode
under a range of values for its amplitude $V$.
It thus also must incorporate the growth of the linear VSI itself. We now examine the various
limits that correspond to different background VSI amplitudes. These
limits should be understood to proceed chronologically as the VSI mode
grows. In almost all of what follows we assume that $q$ is small.

\subsubsection{The small VSI-amplitude limit}

In this limit we set $V=0$ and expect to recover the instability afflicting
the homogeneous background (the VSI itself),
 as discussed in Section \ref{axiVSI}.
If $V=0,$ equation (\ref{aaxistabgov}) becomes
\begin{align}
&\frac{d^2 u_{2\zeta}}{d\zeta^2} +\frac{2{\rm i}qK_\eta\Omega^2}{\sigma^2} 
\frac{d u_{2\zeta}}{d\zeta} 
-\left(K^2_\eta  
-\frac{\Omega^2 K^2_\eta}
{\sigma^2} 
 \right)u_{2\zeta}
=0,\label{aaxistabgovlow}
\end{align}
Because the background VSI has $s \propto |q|,$ we must retain $q$
and the limiting case corresponding to the marginally stable
VSI mode, which has $k_z = -2qk_z.$
We then look for solutions of (\ref{aaxistabgovlow}) with constant coefficients 
such that
$$\u_{2\zeta} \propto
\exp(\text{i}K_\zeta\zeta).$$
This procedure leads to the dispersion relation
\begin{align}
&s^2 = -\sigma^2 =-\Omega^2\frac{K_\eta^2 -2qK_\eta K_\zeta}{K_\eta^2+K_\zeta^2}\label{disbg}
\end{align}
Noting that for $k_z=-2qk_x,$ the Cartesian components
of the wavenumber are related to $K_\eta$ and $K_\zeta$ through
$$ K_x=\frac{ K_\zeta +2q K_\eta }{\sqrt{1+4q^2}}, \quad K_{z}=\frac{K_\eta-2qK_\zeta}{\sqrt{1+4q^2}},$$
then equation (\ref{disbg}) can be writen in the form
\begin{align}
&s^2 = -\sigma^2 =-\Omega^2\frac{K_z^2 +2qK_zK_x}{K_x^2+K_z^2}\label{disbg1}
\end{align}
which is the same as (\ref{growthrate}). Thus the VSI re-emerges as
expected. In this limit the parasites are absent.

\subsubsection{Moderate VSI amplitudes}\label{smallq}

The characteristic time scale associated
with the VSI's shear is $\sim (kV)^{-1}$ and the characteristic growth rate expected 
for a parasitic shear instability is thus $|\sigma| \sim kV$, where
the scale
associated with the mode is assumed to be $\sim k^{-1}\sim K^{-1}_\eta$.
In order for this timescale to dominate that of the background VSI, we require
$|kV|\gg s\sim  |q|\Omega.$ When $q$ is small, as anticipated in real disks,
this leaves a significant range of amplitudes $V$ for which this
regime applies ( $V\gg |q|\Omega/k$). 
Setting $q \rightarrow 0$ in (\ref{aaxistabgov}) yields the governing equation 
\begin{align}
&\frac{d^2 u_{2\zeta}}{d\zeta^2}
-\left(K^2_\eta + \frac{V K_\eta \bar{u}_{\eta}\,(d^2 f/d \zeta^2)}{\overline{\sigma}}
-\frac{2\Omega K^2_\eta\Lambda}
{\overline{\sigma}^2} 
 \right)u_{2\zeta}
=0,\label{aaxistabgovh}
\end{align}
 We remark that in this limit
$$\Lambda = \ez\cdot\boldsymbol{\omega}_\text{vort}= 
\nabla\times(\u_0+\u_1)+2\Omega\ez=V\bar{u}_y df/d\zeta+\frac{1}{2}\Omega,$$
is the vertical component of the
vorticity of the background in the shearing sheet.
The associated term introduces a second order singularity in Eq.~\eqref{aaxistabgovh} 
that is
 stabilising due to the severe wave absorption that occurs at locations 
where $\overline{\sigma}=0$. The physics is analogous to critical
layer formation in the atmosphere and trapped inertial waves in
diskoseismology (Booker and Bretherton 1967, Li et al.~2003).

 When the  VSI mode is localised in $\zeta$ (for instance if it were
 an isolated `wave packet'), 
 it is natural to look for unstable modes that are  also localised in $\zeta.$
In order to proceed further we define $Z= u_{2\zeta}/{\overline{\sigma}}^{1/2}.$
This variable satisfies the equation
\begin{align}
&\frac{d} {d\zeta}\left({\overline \sigma}\frac{d Z} {d\zeta} \right)
-\left(   {\overline \sigma}  K^2_\eta +  \frac{V K_\eta \bar{u}_{\eta}}{2}\frac{d^2 f}{d \zeta^2}
-\frac{{\cal B}}
{\overline{\sigma} } 
 \right)Z
=0.\label{aaxistabgovhh}
\end{align}
where
$${\cal B}=2\Omega K^2_\eta\Lambda -\frac{1}{4} \left(V K_\eta \bar{u}_{\eta}\frac{\partial f }{\partial \zeta}\right)^2.$$
We can find a condition that must be satisfied for a localised unstable mode to exist by multiplying (\ref{aaxistabgovhh}) by $Z^*$
and integrating over the $\zeta$  domain, $D_{\zeta},$  and then
taking the imaginary part. The assumption of localisation enables
boundary terms to be neglected.  
This is also the case for periodic boundary conditions, though care
must be taken when implementing them.

If the parasite is to grow, then $\sigma$ must have a non-zero
imaginary part. This is the case when
\begin{align}
&\int_{D_{\zeta}} \left(\left |\frac{d Z} {d\zeta} \right |^2+ K^2_\eta|Z|^2
+\frac{{\cal B}}
{|{\overline{\sigma} }|^2} 
 |Z|^2\right)d\zeta
=0.\label{aaxistabint}
\end{align}
A necessary condition for a localised unstable mode to exist
is that somewhere in the domain the integrand
changes sign. As the first two terms are positive definite, this
means $\mathcal{B}<0$ somewhere, or rather
\begin{equation}
\Omega^2   +  2\Omega V\bar{u}_y \frac{df}{d\zeta}  
 -\frac{1}{4} \left(V  \bar{u}_{\eta}\frac{d f }{d \zeta}\right)^2 < 0.\label{stabcond}
\end{equation}
It is apparent that rotation, represented by the first term, acts as a
stabilising mechanism. In order to drive instability, the second two
terms must overwhelm the first which happens when $kV
\gtrsim \Omega$, i.e. the characteristic timescale of the VSI's fluid
motions (not its growth rate) must be at least comparable to the
rotation frequency. This can be made somewhat more precise.
Noting that the fastest growing VSI mode possesses
$\bar{u}_\eta=2\bar{u}_y\approx 2/\sqrt{5}$, and finding a $\zeta$ for which
$df/d\xi=-1$ (as is possible for a sinusoid), Eq.~\eqref{stabcond}
 can be expressed as the necessary (though insufficient) instability
 criterion
\begin{equation} \label{precise}
\frac{kV}{\Omega} > \sqrt{5}\left(\sqrt{2}-1\right) \approx 0.9262.
\end{equation}
 Put another way, a necessary
condition for instability is that the Rossby number associated with
the VSI, Ro=$Vk/\Omega$ must be larger than some order-one critical
value. This condition is directly analogous to shear instability in a
stratified medium (the `Richardson criterion'), with Ro functioning like
the inverse Richardson number, and rotation playing the role of
buoyancy. In fact, this wave-damping effect 
can prohibit the existence of discrete normal modes altogether
(see Latter and Balbus 2009).

In summary, we expect 
the advent of parasitic instability to be delayed until the VSI
amplitude ($V$) is rather large. In contrast to the MRI, in
which parasites can always grow but not always sufficiently fast, the
VSI parasites cannot grow at all — not until stabilising rotation is
overwhelmed. In the next subsection we investigate the case of large
VSI amplitude, when this is assured.

\subsubsection{The  large VSI amplitude limit  }\label{largeKb} 

We assume that $|kV|/\Omega \gg 1$ such that
rotation is negligible. In this regime we may neglect the third term in
(\ref{aaxistabgovh}) to obtain
\begin{align}
&\frac{d^2 u_{2\zeta}}{d\zeta^2}
-\left(K^2_\eta + \frac{V K_\eta \bar{u}_{\eta}
(d^2 f/d\zeta^2)}
{\overline{\sigma}}
 \right)u_{2\zeta}
=0,\label{aaxistabgovhhh}
\end{align}
This equation is identical to the Rayleigh equation governing a  plane incompressible shear flow
with velocity profile given by  $v_{\zeta}~=~V \bar{u}_{\eta}f(\xi).$
This is familiar in studies of the Kelvin-Helmholtz instability and its 
properties are well known (see eg. Drazin \& Reid 2004),
 with localised velocity bumps
of the type we consider being generically unstable.
 Key results due to Rayleigh and Howard
 follow on directly. A necessary condition for instability is
that the VSI modes possess an inflexion point in their spatial
structure, i.e. for instability to  occur, we must have $ f'' =0$ somewhere in the
flow.
In addition, an upper bound on the
growth is roughly $|kV|$, and so we obtain
confirmation that parasitic growth does scale with $V$, as assumed earlier.

\subsubsection{Numerical calculations}

To illustrate the behaviour of the axisymmetric parasites for
different $V$, we numerically calculate the
parasitic growth rates for a fiducial VSI mode:
$f(\xi)=\sin\xi$, $q=-0.01$, $k_x/k_z=1/q$. Equations \eqref{axieq1}-\eqref{axieq2} are solved
in the limit of small $q$, so that we may calculate explicit growth rates
$|\sigma|$. The resulting Floquet eigenvalue problem is solved with a pseudo-spectral method (Boyd
2001). 

We examine only modes with zero Floquet exponents, as they are
the fastest growing. For each $V$ we maximise the growth rate over the
wavenumber $K_\eta$ and plot our results in Fig.~4. For large Rossby
numbers, $kV/\Omega$, we
see clearly that $|\sigma|\propto kV/\Omega$, as expected for a shear
instability, and that the growth rate exceeds the rotation rate: the
modes are very fast growing.
The $\eta$-wavenumber of maximum growth is $\approx 0.6k$
generally. As the Rossby number approaches 1, the growth rate drops,
and at the critical value $kV_\text{crit}/\Omega\approx 1.38$ the
Kelvin-Helmholtz parasites die off, stabilised by rotation (consistent
with Eq.~\eqref{precise}). At this
point the algorithm picks up the growth of the VSI itself, with the
low rate $\sim |q|\Omega$. Of course, during the VSI evolution this
procedure occurs in reverse: initially $V$ is small and then slowly
grows to moderate and then to large amplitude.

\begin{figure}
\center
\includegraphics[width=8cm]{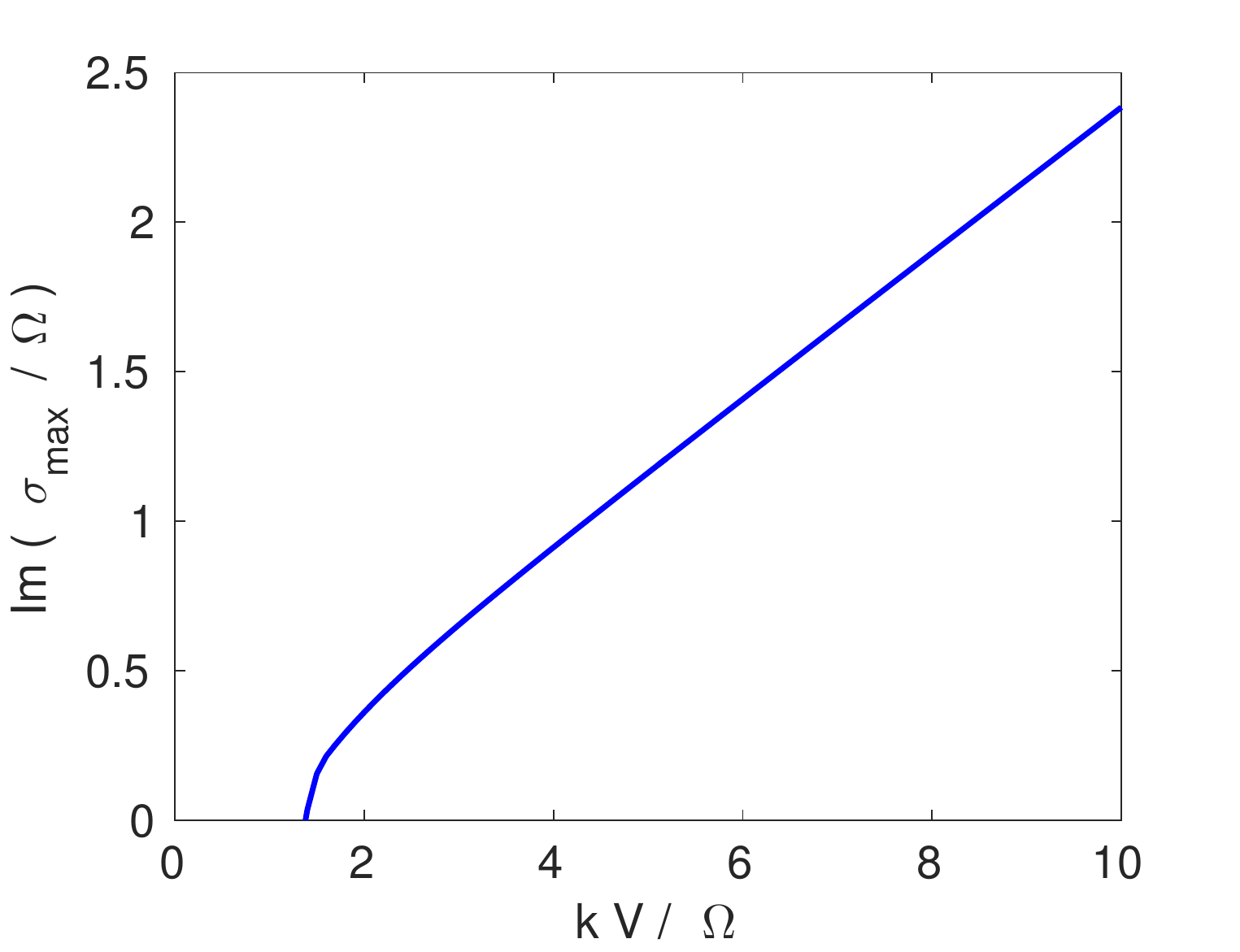}
\caption{Parasitic growth rates maximised over $K_\eta$ for different
  VSI amplitudes (i.e.\ Rossby numbers) $kV/\Omega$. Here $q=-0.01$
  and the underlying VSI mode possesses sinusoidal structure,
  $f(\xi)=\sin\xi$. Because $V$ increases with
  time (at the slow rate $|q|\Omega$) the $x$-axis may be thought of
  as a proxy for time. }
\label{fig::parasitegrowth}
\end{figure}

\subsection{Non-axisymmetric instability }

 We next consider  non-axisymmetric parasitic instabilities. Because
 of their transient growth, such parasites may only be important when the
 VSI achieves
 large amplitudes, but when in this regime they should dominate because
 they can more effectively extract
 shear energy than their axisymmetric counterparts. They are also
 of interest because they are clearly the means by which
 axisymmetry breaks down in the VSI dynamics. The vortices they
 produce will be better aligned with the disk plane, unlike those that
 originate from the
 axisymmetric parasites discussed in the previous subsection.

After writing down the linearised equations, we adopt the same approximation schemes
as in Sections \ref{smallq} and \ref{largeKb} with the aim of deriving equations
analogous to (\ref{aaxistabgovh}) and (\ref{aaxistabgovhhh}).
Assuming the $y$ dependence of modes is such that 
$$ \u_2,\,P_2 \propto \text{exp}(\text{i}K_y y)$$
where $K_y$ is the azimuthal  wavenumber,
we obtain the perturbation equations
\begin{align}\label{nonaxieq1}
\text{i}\overline{\sigma} 
u_{2\zeta} &= -\d_\zeta h_2 + \frac{2\Omega k_x}{k} u_{2y},\\
\text{i}\overline{\sigma} 
 u_{2y} &= -\text{i}K_y h_2 - (V\bar{u}_{1y}\d_\zeta f)\,u_{2\zeta} -
\tfrac{1}{2}\Omega u_{2x} + q\Omega u_{2z}, \label{nonaxiyeq}\\
\text{i}\overline{\sigma}
 u_{2\eta} &= - (V\bar{u}_{1\eta}\d_\zeta
f)\,u_{2\zeta}- \d_\eta h_2 -\frac{2\Omega k_z}{k} u_{2y}, \label{nonaxietaeq}\\
&0= \d_\zeta u_{2\zeta} + \d_\eta u_{2\eta} + \text{i}K_y u_{2y},\label{nonaxieq2}
\end{align}
where $h_2=P_2/\rho_0$, as before, and
$$ \overline{\sigma}=-\text{i}\d_t -K_y(3x/2+ qz)\Omega + 
V(\mathbf{K}\cdot\bar{\u}_1)\,f(\xi),$$
with $\mathbf{K}= K_y\ey-   {\rm i} \d_{\eta}\mathbf{e}_\eta$.
We remark that in the new coordinates
$$ \tfrac{3}{2}x+qz=(\tfrac{3}{2}k_x+qk_z)\zeta/k+(qk_x-\tfrac{3}{2}k_z)\eta/k.$$

On account of its explicit dependence on $\zeta, \eta$ and $t,$ the above system is
not strictly separable. But if we employ the small $|q|$ limit as in  Section \ref{smallq},
some separability may be
restored. 
The fastest growing VSI modes possess
$k_z  = O(qk_x),$ and so
 $$ \tfrac{3}{2}x+qz=\tfrac{3}{2}\zeta+O(q\eta).$$
 Thus in the limit $|q| \rightarrow 0,$ the system becomes separable in $\eta$ and $t$
 as well as $y$ so that we may assume that
 $$ \u_2,\,P_2 \propto \text{exp}(\text{i}(\sigma t+ K_yy+K_\eta\eta).$$
 Then
 \begin{equation}\overline{\sigma}\rightarrow \sigma  -  \frac{3}{2}K_y\zeta \Omega + V(\mathbf{K}\cdot\bar{\u})\,f(\xi), \label{shearM}\end{equation} 
and $\mathbf{K}  \rightarrow  K_y\ey+   K_\eta\mathbf{e}_\eta. $
A second order ODE may now be derived which we consider
in two limits.

\subsubsection{Tight-winding limit}

 The first limit we treat is the tight winding approximation for which the Keplerian 
shear may be neglected. From (\ref{shearM}) this requires
 $K_y\Omega/k   \ll  |{\bf K} V|.$
 Assuming, as above, that $|{\bf K}|/k\sim 1,$ this corresponds to
 $K_y/K_\eta   \ll  |{\bf K}V |/\Omega.$ Thus in the limit of sufficiently small $K_y/K_\eta,$ the Keplerian shear may be neglected.
 In this `tight-winding limit' the set  of equations (\ref{nonaxieq1}) -  (\ref{nonaxieq2})
becomes equivalent to the set  (\ref{axieq1}) -  (\ref{axieq2}) that governs the axisymmetric problem.
Accordingly, equation (\ref{aaxistabgovh}) is recovered in the small $q,$ small $K_y/K_\eta$ limit.

\subsubsection{The  large VSI-amplitude limit}\label{largeKbna}  

 In this limit one assumes, in addition to small $|q|,$  
  that  $|kV|/\Omega \gg 1$, but  there is no requirement that $K_y$ be small.
 In this case all the terms explicitly proportional to $\Omega$  in equations (\ref{nonaxieq1}) -  (\ref{nonaxieq2})
can be dropped.
The system reduces to a single second-order ODE
\begin{equation}\label{nonaxiDE}
\frac{d^2 u_{2\zeta}}{d\zeta^2} -\left(K^2 +
 \frac{ V(\mathbf{K}\cdot\bar{\u})(d^2f/d\zeta^2)}{\sigma  + 
V(\mathbf{K}\cdot\bar{\u})\,f(\xi)}\right)
u_{2\zeta} = 0.
\end{equation}
This is again a form
of Rayleigh's equation and the
 discussion  in Section \ref{largeKb}  
regarding (\ref{aaxistabgovhhh})
applies here also. Note that the non-axisymmetric growth rates will be
larger than the corresponding axisymmetric growth rates in this limit
by a factor $K/(K_\eta \overline{u}_\eta) \approx 2$. This is because
the non-axisymmetric wavevector is free to be parallel to
the direction of the shear flow. 

\subsection{Discussion}

Putting together these theoretical details, we can construct a
relatively straightforward account of a VSI mode's growth and
evolution. 

First, the VSI will grow relatively unhindered up to
`moderate'
amplitudes, $0< V < \Omega/k$. It is a nonlinear solution, so there are no nonlinear
effects to intervene. During this time,
nonaxisymmetric parasitic modes will fail
to get any traction because they only grow transiently over
a shear time $1/\Omega$: if their growth rates during this time are (at
most) proportional to $kV$ then they will only amplify by a factor
$\text{exp}[(Vk/\Omega)(\text{e}-1)]\lesssim 5.575$, most likely
insufficient to overtake their host.
Axisymmetric parasites are untroubled by the shear; 
however, they are strongly stabilised by the disk's
rotation when at these low Rossby numbers (as explained in Section
3.1.2 and illustrated in Fig.~4). Only when the VSI amplitude is above a critical value
$V>V_\text{crit}\gtrsim \Omega/k$ will axisymmetric parasites grow
exponentially, and ultimately disturb the underlying mode. 
Near the same point nonaxisymmetric parasites may also achieve greater amplitudes during their windows of growth. This will
all occur at some second critical value $V=V_\text{max}$. It is unclear
whether the axisymmetric or non-axisymmetric parasites will
launch the first successful attack. It is possible that they may emerge
concurrently.

\subsubsection{Maximum VSI amplitudes}

The precise value of $V_\text{crit}$ is determined by the details of
the flow, in particular by the form of $f(\xi)$. The value of the
maximum VSI amplitude $V_\text{max}$ is the amplitude at which the VSI mode is
significantly disrupted. It is necessarily larger than $V_\text{crit}$, and
 may be estimated using arguments
presented in Latter (2016). 

Suppose the VSI grows according to the
equation $V(t)= V_\text{crit} \,\text{exp}(|q|\Omega t)$, so that at
$t=0$ the VSI has hit the first critical amplitude.
Suppose from that point the parasite amplitude $p$ grows according to 
$dp/dt = kV(t)p(t)$, with $p=p_0$ at $t=0$. The latter ODE can be
solved and the time at which the parasite amplitude equals its host
($p=V$) calculated, an event that 
occurs when the
VSI attains the amplitude:
\begin{align}
V = V_\text{max} &=
-\frac{q\Omega}{k}W_{-1}\left[-\frac{kp_0}{q\Omega}\text{exp}\left(\frac{-k
        V_\text{crit}}{q\Omega}\right)\right]  \\
   &\gtrsim V_\text{crit}\left[1+
  |q|\ln\left(\frac{|q|}{kp_0/\Omega}\right)  \right],
\end{align}
where $W_{-1}$ is the second branch of Lambert's function,
and the final estimate comes from its asymptotic expansion in small
argument and by setting $V_\text{crit}\gtrsim \Omega/k$ (see Corless et
al.~1996, Latter 2016). The second term in the square brackets
depends on the relative sizes of $|q|$ and the noise
from which the parasite grows. Even if this ratio is large, it will be
the moderated by the log and the $q$ prefactor, so for simplicity we set
$V_\text{max}\gtrsim V_\text{crit} \gtrsim \Omega/k$. 
In summary, we do not expect these three velocity scales to differ by very
much.

In order make to develop the theory further and make concrete estimates, 
we need to know what value $k$
takes, i.e.\ the characteristic wavenumber of the fastest growing VSI mode,
which means invoking physics so far neglected.
 Consider a disk region
in which the diffusion approximation holds (at 1 AU, say, in the 
low-mass disk
model of Lesur and Latter 2016). From Section 2.4 and Fig.~3, we
observe that the fastest growing VSI modes possess
$k>\sqrt{\Omega/\kappa}$. An upper bound on $V_\text{max}$ is hence
$$V_\text{max}\sim \sqrt{\kappa\Omega}\sim \text{Pe}^{-1/2}c_s,$$
where $c_s$ is the sound speed and Pe is the Peclet number, as
earlier. At 1 AU in our model Pe$\sim 10^3$, and so the characteristic 
VSI amplitudes are a few percent of the disk sound speed. More massive
disk models may give a similar estimate out to radii as far as 10 AU. 
In our low-mass model, however, larger radii are only marginally within
the compass of the local approximation, or are outside the diffusion
approximation. Nonetheless, 
the estimated velocity amplitudes might increase to some 10\% of the
local sound speed, and hence should be significant. 

An alternative, and more general,
upper limit on $V$ can be obtained by noting
that $k\sim k_x \sim |1/q|k_z \gtrsim |1/qH|$, where $H$ is the disk
scaleheight. VSI modes with $k_z\sim 1/H$ might correspond to the
`body modes' witnessed in global simulations and vertically stratified
boxes (Nelson et al.~2013, Barker and Latter 2015). This then implies
that $V_\text{max} \lesssim |q| c_s \sim (H/R) c_s$. Once again, we
find that the maximum turbulent speeds should be a few percent to 10
percent the local sound speed. 

\subsubsection{Saturation and comparison with simulations}

These maximum amplitudes we expect to be attained at the
beginning of the VSI evolution. 
But what happens next, after the system
breaks down into turbulent state and saturates? What velocity
amplitudes might we expect in this state? The key to this lies in
the radiative driving of the vertical shear and its relative strength
in comparison with the VSI. One can envisage two limits.

First, suppose that the vertical shear is strictly enforced by the star's
radiation field, and so the VSI is relatively ineffective in smoothing
it out. This is certainly the scenario in our shearing box model, in
which the vertical shear is `hard-wired' into the box. It is also the case for
isothermal global simulations or ideal gas 
simulations with short relaxation times (Nelson et al.~2013 ).
The VSI will churn away upon the vertical shear, but be unable
to erase it.
One might then be tempted to ascribe the general VSI saturation to the
parasitic modes, rather than a weakening of the underlying unstable
state. The parasites limit VSI growth via a transfer of
energy to smaller scales and ultimately the viscous length, and will
set their characteristic amplitude. This turbulent amplitude 
would then be $\lesssim V_\text{max}$, a few percent or more of the sound speed,
and indeed this is in rough agreement with global simulations (Nelson et
al.~2013, Stoll and Kley 2014).
In fact, the relatively large
amplitudes produced by these simulations, and the fact that the saturated state is
characterised by nonlinear structures similar in form to linear modes,
provides strong evidence in favour of this interpretation. Large amplitudes come about
because (a) the linear modes are nonlinear solutions and (b)
parasitic modes that might limit them
are hampered by the stabilising effect of rotation. 

Parasitic modes in this context may not only limit the amplitudes
of the VSI but also direct energy from them into the formation of nonaxisymmetric
structures, such as vortices. Being essentially Kelvin-Helmholtz in
nature, the `wrapping up' of the VSI vortex layers is a natural outcome of
their evolution. As shown in Richard et al.~(2016), vortex formation
is key to the breakdown of axisymmetry in the VSI saturation, and
hence to the possibility of any accretion. We observe that the
vortices formed in our local model are not oriented flush with the
orbital plane, but are aligned with the VSI flow $\bar{\u}$. 
Consequently, they are
tilted upward or downward relative to the orbital plane by an angle
$\approx 60^\circ$, and will not appear columnar as in
 the
subcritical baroclinic instability or `Rossby wave' instability (Lesur
and Papaloizou 2010, Lovelace et al.~1999). Indeed, Richard et
al.~(2016) find that the vortex structures possess both radial and
vertical variations, on some length of order $0.1 H$. According to our analysis
this wavelength should correspond to that of the fastest
growing parasites, which in turn will approximate the radial wavelength of
the underlying VSI modes. This is also in keeping with
the global simulations. 

The second limit corresponds to a scenario in which
 the baroclinic
driving is very weak and the VSI overpowers the vertical shear,
effectively smoothing it out. Initially the VSI might achieve large
amplitudes but, as the destabilising gradient (the vertical shear)
is destroyed, it
will settle down into a low-amplitude sluggish state near marginal
stability (as in certain simulations in Stoll and Kley 2014). 
In this case the saturated amplitude will be $\ll
V_\text{max}$. Any vortices, or other non-axisymmetric structure,
created in the initial burst will ultimately die and axisymmetry
will be restored. Consequently, no accretion is to be expected. 

In reality, most disks at a given time will lie somewhere in between these two
extremes. Detailed modelling of the baroclinic driving and a sequence
of careful and detailed global simulations are
needed to determine the expected range of states the VSI occupies.
It should be stressed that in order to sustain vortex production (and accretion) the VSI must 
be permitted to achieve large amplitudes. If it is too efficient, then
it will settle down into a less interesting low state.

\section{Vertical shear and magnetic fields}

In certain regions of a protoplanetary disk magnetic fields may 
have some influence, even in dead zones where the MRI is suppressed. 
Certainly in the upper layers of disks, which are subject to
significant photoionisation,
the gas may couple effectively to a mean magnetic field, which in turn may be too
strong to permit the MRI, but rather a magnetocentrifugal wind (Bai
and Stone 2013, Lesur et al.~2014, Bai 2014).
We may then ask what effect does magnetism have on the onset of the VSI? 
Conversely, MRI-active regions may exhibit vertical shear:
how would that impact on the MRI?

In this section we make a start on these questions by determining the
linear response of ionised fluid pierced by a mean magnetic field.
Non-ideal effects are only dealt with in passing, and could form the
basis of future work.

\subsection{Governing equations and magnetic equilibria}

The equations governing our ionised local slab of disk are now
\begin{align} \label{mhd1}
\d_t\u + \u\cdot\nabla\u &= -\frac{1}{\rho}\nabla P_\text{tot}
-2\Omega\ez\times\u  \notag \\
&   \hskip1.5cm -\Omega (3x  +2z q ) \ex +\frac{\B\cdot\nabla\B}{4\pi\rho} , \\
\d_t\B + \u\cdot\nabla\B &= \B\cdot\nabla\u,
\end{align}
alongside $\nabla\cdot\u=\nabla\cdot\B=0$, where $\B$ is the magnetic
field, and $P_\text{tot}$ denotes the sum of gas and magnetic
pressures. 

This system admits the following equilibrium, similar to before
$$ \u=\u_0=-\frac{1}{2}(3x+2qz)\ey,\quad  P_\text{tot}=P_0,\quad
\B=\B_0,$$
where $P_0$ is a constant scalar and $\B_0$ is a constant vector.
We are completely free to specify the  $y$ component of $\B_0$
but the radial and vertical components  are constrained by Ferraro's law, i.e. the $y$
component of the induction equation $\B_0\cdot\nabla\u_0=0$. This 
states that in order to have a steady equilibrium we must have
\begin{equation}
B_{0x}= -\frac{2q}{3} B_{0z}.
\end{equation}
In other words, the $B_y$ generated by the radial shear (working on
$B_x$) must be exactly balanced by the $B_y$ created by the vertical
shear (working on $B_z$). This is a key issue in constructing magnetic
equilibria in the presence of vertical shear: both vertical and radial
magnetic fields must be present in the correct amounts. 
(This issue was overlooked in Urpin and Brandenburg 1997.) Global
examples of such fields have been computed by Ogilvie (1997),
  and vertically stratified examples by Riols et al.~(2016) (see also
  Papaloizou \& Szuszkiewicz, 1992). Typically
 the poloidal field varies with space but in our simple local  incompressible
 model the magnetic field is conveniently  constant.
Finally,  for simplicity, we set the $y$ component of $\B$ to
zero. 

\subsection{Perturbations and the
  dispersion relation}

To understand the two instabilities in question we perturb the
equilibrium described above by disturbances $\u'$, $P_\text{tot}',$
and $\B'$. Assuming that they are $\propto \text{exp}(\text{i}k_x x +
\text{i}k_z z + s t)$, the linearised equations governing their evolution may be
written as
\begin{align}
s u_x' &= -\text{i}k_x h' + 2\Omega u_y' + \text{i} 
(\mathbf{v}_A\cdot\mathbf{k}) b_x', \\
s u_y' &= -\tfrac{1}{2}\Omega u_x' + \Omega q u_z' + \text{i}
(\mathbf{v}_A\cdot\mathbf{k})
b_y', \\
s u_z' &= -\text{i}k_z h' + \text{i} (\mathbf{v}_A\cdot\mathbf{k})  b_z',\\
s b_x' &= \text{i}(\mathbf{v}_A\cdot\mathbf{k})  u_x', \\
s b_y' &= \text{i} (\mathbf{v}_A\cdot\mathbf{k})  u_y' - \tfrac{3}{2}\Omega b_x' - q\Omega
b_z', \\
s b_z' &= \text{i}(\mathbf{v}_A\cdot\mathbf{k})  u_z', \\
0&= \text{i}k_x u_x' + \text{i}k_z u_z',
\end{align} 
where $h'= P_\text{tot}'/\rho$ with the perturbed and background 
Alfv\'en velocities
$\mathbf{b}'=\B'/\sqrt{4\pi\rho}$ and
$\mathbf{v}_A=\B_0/\sqrt{4\pi\rho}$, respectively, and $\mathbf{k}=\ex
k_x+ \ez k_z$.
 On account of the incompressibility condition the solution automatically satisfies $\nabla\cdot\B'=0.$
 This means that it satisfies the nonlinear equations as well as the linear ones.
But note that when $k_x=0$, incompressibility enforces $u_z'=0$ and
consequently $b'_z=h'=0$. In that case terms involving $q$
disappear, and the resulting MRI `channel flows' do not feel the vertical
shear whatsoever. 

After some algebra these equations can be reduced to a
biquadratic dispersion relation for the growth rate $s$:
\begin{align}
& s^4 + \left[  2(\mathbf{v}_A\cdot\mathbf{k})^2+\epsilon^2\Omega^2
     +2\epsilon^2\Omega^2q \frac{k_x}{k_z}   \right]s^2 \notag\\
& \hskip0.4cm+ (\mathbf{v}_A\cdot\mathbf{k})^2\left[
 (\mathbf{v}_A\cdot\mathbf{k})^2 -3\Omega^2\epsilon^2 + 
  2\epsilon^2\Omega^2q \frac{k_x}{k_z} \right]=0, \label{MHDdisp}
\end{align}
in which we have
introduced the additional notation
$$ \epsilon\equiv \frac{k_z}{k}=\left(1+\frac{k_x^2}{k_z^2}\right)^{-1/2},$$
as in Latter et al.~(2015). 

We can obtain  stability criteria by first noting that the two roots for $s^2$ 
obtained from (\ref{MHDdisp})  are always real. 
Instability occurs when
one of these is positive, which happens when either (a)
 the coefficient of $s^2$ is negative
or (b)  when this coefficient is positive but the
last term in the dispersion relation is negative.
When there is no background magnetic field
($\mathbf{v}_A=\mathbf{0}$),
 the condition (a) reduces to the hydrodynamical 
VSI instability criterion, Eq.~\eqref{hydrostab}, which 
states that $q(k_x/k_z)<-1/2$.  
The presence of
a non-zero magnetic field
($\mathbf{v}_A\neq\mathbf{0}$), no matter how small, changes the
picture abruptly. Then 
the condition (b) is the easiest of the two to satisfy, yielding
\begin{equation}\label{stab}
 (\mathbf{v}_A\cdot\mathbf{k})^2 < \frac{k_z^2}{k^2}\Omega^2
       \left(3 - 2q\frac{k_x}{k_z}\right),
\end{equation}
that is, when the Alfv{\`e}n frequency is less than the characteristic
frequency defined by the right hand side. This criterion captures both
the MRI and the VSI.

\subsection{The MRI and vertical shear}

We first examine how the MRI is altered by
vertical shear. When $q=0$, we reproduce the well known results for
the purely vertical field MRI, $\B_0\propto \ez$. The fastest growing modes are channel modes,
while `radial modes', with $k_x\neq 0$,  
grow a factor $\epsilon$ slower.
 As explained in
Latter et al.~(2015), radial modes exhibit vertical circulation that
impedes the MRI instability mechanism. An alternative way to think
about this is in terms of the competition between  the effects of rotation
and magnetism:
when $k_x=0$ the characteristic
rate of destabilisation, on account of the rotation profile, 
is $\propto \Omega$, but when $k_x\neq 0$
it is $\propto \epsilon\Omega<\Omega$. The stabilising influence of 
magnetic tension is accordingly more effective for a given Alfv{\`e}n frequency
$(\mathbf{v}_A\cdot\mathbf{k})$.
 As a consequence of this, we expect the MRI branch of modes to be most
important (and the most characteristic) for wavenumbers oriented nearly
vertically, i.e. when $k_x \ll k_z$, which is in convenient contrast
to the VSI, for which the opposite limit pertains. 

 We find the expression for the squared
 growth rate,  solving  eq.~\eqref{MHDdisp} 
under the assumption that
criterion (b) applies, Eq.~\eqref{stab}. We next consider it as a
function of $(\mathbf{v}_A\cdot\mathbf{k})^2$, and find its maximum
value. 
This occurs when
\begin{equation}\label{critk}
(\mathbf{v}_A\cdot\mathbf{k})^2= \frac{15}{16}\epsilon^2\Omega^2 
-\frac{1}{4}q\frac{k_x}{k_z}\epsilon^2\Omega^2\left(1 + q\frac{k_x}{k_z}\right),
\end{equation}
the first term on the right should be familiar from previous studies
of the MRI, while the second comes from the vertical shear.
 Substituting this expression into the growth rate obtained from
 \eqref{MHDdisp},
 we obtain a
remarkably simple
expression for the maximum rate:
\begin{equation}\label{maxs}
s_\text{max}= \frac{3}{4}\epsilon\Omega\left(1-\frac{2}{3}\frac{k_x}{k_z}q\right).
\end{equation}
It corresponds to the classical expression for maximum growth rate  for the
MRI (e.g.\ Latter et al.~2015) with a `correction' proportional to $q$
that comes from the vertical shear. The bracketed expression in Eq.~\eqref{maxs} is, in fact, proportional
to $(\mathbf{k}\times(\nabla\Omega))\cdot {\bf e}_y$, and so complete stabilisation
occurs for modes with $\mathbf{k}$ parallel to the angular velocity
gradient. The components of the wavenumber then satisfy  $k_x/k_z= 2/(3q)$.

 Furthermore, we may maximise
\eqref{maxs} with respect to $k_x/k_z$, and find that $s_\text{max}=
(3/4)\Omega + (5/81)q^2\Omega +\mathcal{O}(q^4)$, assuming small $q$,
and this occurs when $k_x/k_z\approx -(1/9)q$. Essentially this is a
channel mode, but the slightly elevated growth rate
indicates that the MRI can draw some
energy from the background vertical shear in addition to the radial
shear, thus acting partly like the VSI.  
Overall, however, the fastest growing and most important
MRI modes, for which $k_z>k_x$, are not especially impacted by the vertical shear, if indeed
we expect $q\ll 1$. 

\subsection{The VSI and magnetic fields}

The VSI lies on the same branch of
the dispersion relation as the
the MRI, and so it is not easy to distinguish the two: as $k_x/k_z$
goes from very small to very large values the MRI smoothly morphs into
the VSI. To illustrate this point we plot in Fig.~5 growth rate contours in the 
 $[(\mathbf{v}_A\cdot\mathbf{k})/\Omega_0,\, k_x/k_z]$ plane for $q=0$ and
 $q=-0.3$. In the latter case we observe low growth for large $k_x/k_z$
and small $(\mathbf{v}_A\cdot\mathbf{k})/\Omega_0$,  which we associate with the
VSI, and large growth for small $k_x/k_z$ and 
$(\mathbf{v}_A\cdot\mathbf{k})/\Omega_0\sim 1$, which we associate with
the MRI.

To disentangle the VSI algebraically, 
we let $k_x/k_z=-1/q$, the wavevector
orientation that yields
maximum VSI growth in the absence of a magnetic field. 
Both expressions \eqref{critk} and \eqref{maxs} should hold for such a
mode. Plugging in our value for $k_x/k_z$ we find that maximum growth
occurs at very small values of $(\mathbf{v}_A\cdot\mathbf{k})^2\sim
q^2$, but with maximum growth $s_\text{max} = (5/4)q\Omega$, slightly
larger than in the hydrodynamical case. This indicates that for exceedingly
light magnetic tension, the VSI is slightly amplified.

Overall, however, the impact of magnetic tension is stabilising,
with stability occurring when \eqref{stab} is violated.
For VSI wavevector orientations $k_x/k_z\sim -1/q$, the criterion can
be reframed in terms of the vertical component of the magnetic
tension. Unless $v_{Az}k_z$ is very small, the mode is suppressed:
\begin{equation}
v_{Az}k_z \lesssim q\Omega.
\end{equation}
Put another way, for a given $v_{Az}$, the critical vertical
wavelength for instability is long, so as to best escape the
stabilising magnetic tension.

But in order for the VSI modes to even fit into the disk these
lengthscales must be less than $H$. This
furnishes us with an instability criterion. We find that
the VSI can only occur in magnetised disks if
\begin{equation}
\beta \gtrsim q^{-2},
\end{equation}
where $\beta$ is the average plasma beta within $|Z|<H$ (the MRI only requires that
$\beta\gtrsim 1$). Using the estimate $q\sim H/R\sim 0.05$ the criterion
becomes $\beta \gtrsim 400$. Note that the critical $\beta$ may be significantly
larger for smaller scale VSI modes, such as those that typically appear in the surface layers of the 
disk, and furthermore the local $\beta$ in these layers may be low indeed.

\subsection{Ohmic diffusion}

In the dead zones of PP disks, magnetic diffusion will alter
the results of the pervious sections. In particular, it will weaken magnetic tension and the
VSI will find it easier to grow. We now estimate how much diffusion is needed to `rescue’
the VSI.

By comparing frequencies, magnetic diffusion
dominates tension when $\eta k^2 \gtrsim v_A k$, where $\eta$ is
Ohmic resistivity. For fixed $v_A$, this means that on scales $k \gtrsim k_D \equiv
v_A/\eta$ magnetic tension may be overcome and hence neglected.
On the other hand, the VSI is suppressed when its growth rate $\sim
q\Omega$ is equal or less the Alfven frequency $v_Ak$. This occurs on wavenumbers $k\gtrsim
k_S\equiv q\Omega/v_A$. Putting these two estimates together, we
recognise that the VSI is impeded when $k_S \lesssim k \lesssim k_D$, and unimpeded
for other wavenumbers. In fact, magnetic tension is completely subdued by
diffusion for \emph{all} modes when $k_D \lesssim k_S$ which gives the condition
\begin{equation}\label{Elss}
E_\eta \lesssim q,
\end{equation}  
where the Ohmic Elsasser number is defined to be $E_\eta=
v_A^2/(\Omega\eta)$. Note that violation of Eq.~\eqref{Elss} does not
mean the VSI fails to appear: unstable VSI modes will still operate on
sufficiently short (and possibly sufficiently long) scales.

To estimate $E_\eta$ requires knowledge of the strength of the
background vertical field. For a midplane $\beta=10^5$, Lesur et
al.~(2014),
using the minimum-mass solar nebula,
estimate that $E_\eta <10^{-4}$ when $|z|<H$ at $R=1$ AU, and
$E_\eta=0.1-1$ when $R=10$ AU. Given that $q\sim H/R\sim 0.05$, we
recognise that at 1 AU the VSI is free of magnetic tension; but 
further out in the disk it may be worth considering the
role of magnetic fields a little more closely, especially when those fields
are strong.

\begin{figure}
\center
\includegraphics[width=8cm]{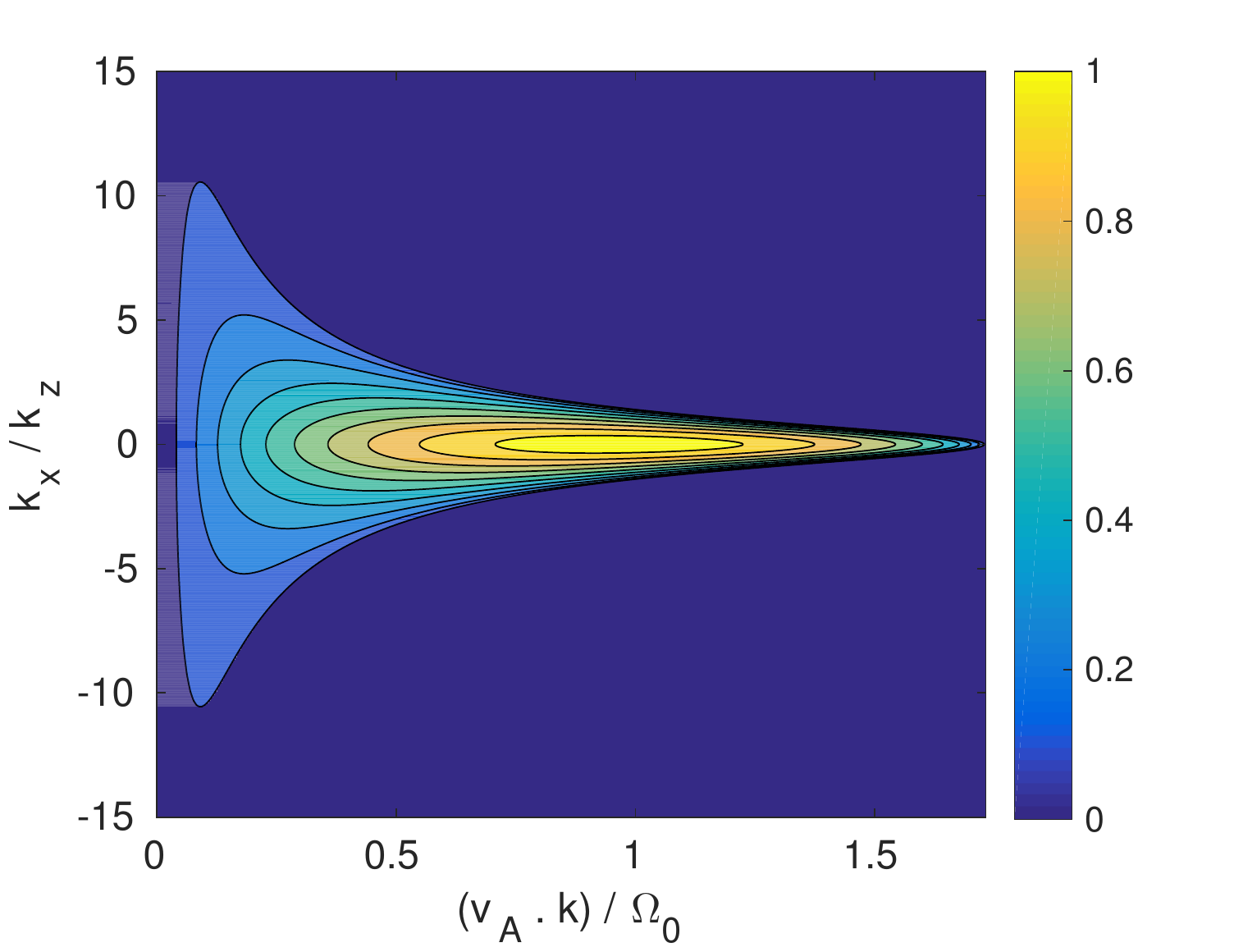}
\includegraphics[width=8cm]{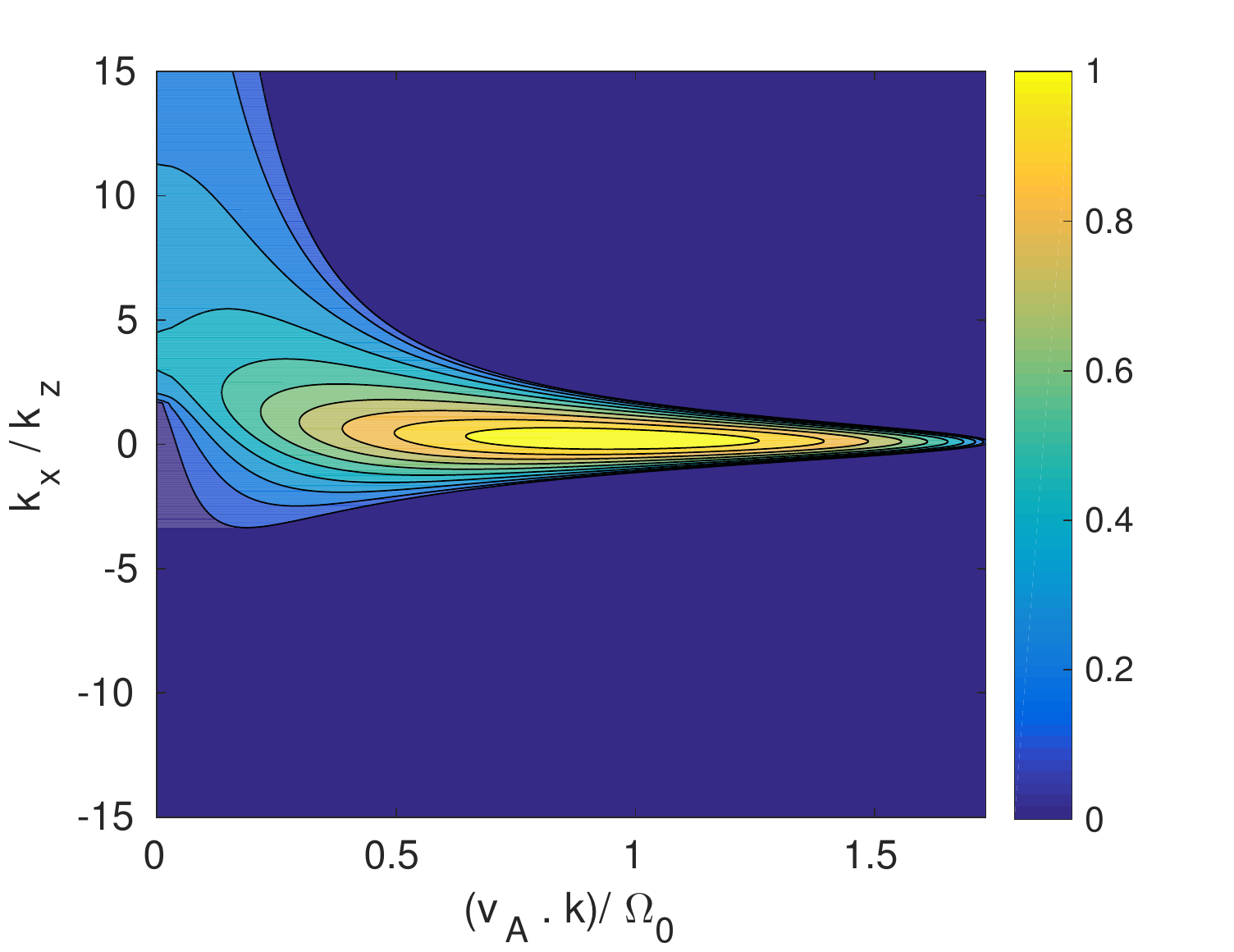}
\caption{Top panel: coloured contours of the MRI growth rate when
  $q=0$ in the
  $[(\mathbf{v}_A\cdot\mathbf{k})/\Omega_0,\, k_x/k_z]$ plane. Note the symmetry
  about the horizontal axis. Bottom panel: contours of the MRI and VSI growth
rates when $q=-0.3$. Observe the marked asymmetry about the horizontal axis,
and the extension of growth to large $k_x/k_z$. }
\label{fig::MHDandshear}
\end{figure}

\section{Conclusion}

In this paper we have established a number of theoretical results
pertaining to the onset and saturation of the vertical shear
instability (VSI) in protoplanetary disks. Using the Boussinesq
approximation, we show that the linear VSI modes are nonlinear
solutions whose spatial structure need not be limited to sinusoids.
Being double-diffusive, 
the instability grows at its maximum rate $\sim (H/R)\Omega$ on 
a range of wavelengths bracketed from below by
$(H/R)^{-1/2}\text{Re}^{-1/2} H$ and above by
$(H/R)^{1/2}\text{Pe}^{-1/2}H$, where Re and Pe are the Reynolds and
Peclet numbers. On sufficiently short scales and in certain disc
regions the diffusive approximation breaks down and these estimates
require moderate revision. In this case, maximum growth is assured if
the gas's cooling rate is much greater than $(R/H)\Omega$ (in
agreement with Lin and Youdin 2015). 
We apply these estimates to a low-mass disk and
demonstrate that the VSI is prevalent from 1 AU outward, moving from
shorter to longer scales with disk radius.  

The VSI modes cannot grow indefinitely, as they are 
subject to parasitic instabilities of
Kelvin-Helmholtz type. The onset of the parasites, however, is
significantly delayed:
axisymmetric instability is impeded by the gas's stabilising 
radial angular momentum gradient, whereas non-axisymmetric instability
is foiled by the disk's shear. As a consequence, the VSI achieves
relatively large amplitudes before breaking down, these characterised
by Rossby numbers greater than 1 and fluid velocities a few percent
or more of the sound speed. This makes a striking contrast to the
convective overstability, whose amplitudes are kept generally low by
parasites (Latter 2016). The delay in VSI disruption might explain some features witnessed
in global simulations, such as the dominance of large amplitude linear
waves, at least initially (Nelson et al.~2013). We note
that our analysis strictly holds only for nonlinear modes that remain
shortscale, and that the body modes are not represented in a
Boussinesq model. Nonetheless, the physical effects outlined above
will also work on larger scales and 
should get in the way of their disruption as well.

The parasites may play an influential part in the VSI's subsequent
saturation. If the background vertical shear
is forcibly maintained by stellar irradiation, the parasites may 
set the amplitude of the resulting quasi-steady turbulent state. 
If, however, the VSI is efficient in erasing the vertical shear, they
may be less important.
Significantly, parasitic instabilities are the route by which the VSI's
axisymmetry is broken; the subsequent turbulence may then transport angular
momentum.
In so doing they create vortices, not perfectly aligned with the disk plane. 
Vortex production is a robust process, and VSI modes on all scales can
generate them. As discussed at length elsewhere, vortices can
accumulate solid particles,
though in this context it may be complicated by their non-trivial vertical structure. 
Their geometry and limited lifetime are other variables deserving of
further study (Richard et
al.~2016). 

Finally we include magnetic fields in the analysis. Because of the
vertical shear it is important to be careful to set up a
meaningful time-independent magnetic equilibrium. We find that the MRI
is not impacted greatly by the vertical shear, and the fastest growing modes (channel
modes) are not affected at  all. In contrast, the VSI is completely suppressed by magnetic
tension, for (average) plasma betas below $\gtrsim q^{-2}\sim
(R/H)^2$. Ohmic diffusion can rescue the VSI, however, if the
local Elsasser number is less than $q\sim H/R$. 

\section*{Acknowledgements}

The authors would like to thank Andrew Youdin, Adrian Barker, and the anonymous referee
for useful comments that helped improve the
manuscript. Special thanks is also due Tobias Heinemann for a close
reading of an earlier draft. This work is partially funded through
STFC grant ST/L000636/1.

\begin{appendix}

\section{The VSI with stratification, cooling, and viscosity}

In this appendix we tackle the complete problem, with viscosity,
vertical buoyancy, and radiative cooling in the framework of the
Boussinesq approximation. Similar analyses have appeared in Urpin and
Brandenburg (1997) and Urpin (2003), though the key results deserve
further clarification.

The governing equations are now
\begin{align}
\d_t\u + \u\cdot\nabla\u &= -\frac{1}{\rho}\nabla P -2\Omega\ez\times\u +
                 \Omega^2 (3x  +2z q ) \ex \notag\\
& \hskip3cm  -N^2\theta \ez +\nu\nabla^2\u, \\
\d_t\theta + \u\cdot\nabla\theta &= u_z + \kappa\nabla^2\theta,
\end{align}
with $\nabla\cdot\u=0$, where $\theta$ is the buoyancy variable, $N^2$
is the squared buoyancy frequency, $\nu$ is the kinematic viscosity,
and $\kappa$ is the thermal
diffusivity, not to be confused with the epicyclic frequency. 

These equations admit the same steady state as appearing in Subsection
2.2 if $\theta=0$. We perturb this equilibrium with perturbations as
earlier, with the perturbed buoyancy $\theta'= \bar{\theta}(t)f(\xi)$,
and we find that such disturbances remain nonlinear
solutions. However, the form of $f$ is constrained by the Laplacians,
so that $f \propto d^2f/d\xi^2$, and so only sinusoidally varying shearing waves are
supported.

We consider only axisymmetric disturbances and assume that they are
the real parts $\propto \text{exp}(st+\text{i}k_x x + \text{i}k_z z)$. We then have
the equations
\begin{align}
s\bar{u}_x &= -\text{i}k_x \bar{h} + 2\Omega \bar{u}_y -\nu k^2
\bar{u}_x, \\
s\bar{u}_y &= -\tfrac{1}{2}\Omega \bar{u}_x + q\Omega \bar{u}_z-\nu
k^2 \bar{u}_y, \\
s\bar{u}_z &= -\text{i}k_z \bar{h} -N^2\bar{\theta} -\nu k^2
\bar{u}_z, \\
s\bar{\theta} &= \bar{u}_z - \kappa k^2 \bar{\theta},
\end{align}
with $k_x\bar{u}_x + k_z \bar{u}_z=0$, $k^2=k_x^2+k_z^2$, and where we have defined
$\bar{h}= \bar{P}/\rho$. Solvability of this set gives us the
dispersion relation
\begin{equation}\label{fullbeast}
s_\nu^2 = s_\text{VSI}^2 -\frac{k_x^2}{k^2}N^2\frac{s_\nu}{s_\kappa},
\end{equation}
in which $s_\text{VSI}$ is the VSI growth rate in the absence of
viscosity, buoyancy, and thermal diffusion:
$$ s_\text{VSI}^2=
-\Omega^2\frac{k_z^2}{k^2}\left(1+2q\frac{k_x}{k_z}\right),$$
and $s_\nu= s+ \nu k^2$ and $s_\kappa = s + \kappa k^2$. If we had
used Newtonian cooling the dispersion relation would only altered by a
redefinition of $s_\kappa$ which would become $s_\kappa= s + 1/\tau$,
where $\tau$ is the cooling rate.

We now analyse Eq.~\eqref{fullbeast} paying attention to different
scales. Suppose first that we are interested in wavelengths much
longer than the viscous length, so that 
\begin{equation}\label{short}
k^2 \ll q(\Omega/\nu).
\end{equation}
 This
means straightaway that $s_\nu \approx s$ for the fastest growing VSI
mode. We then examine on what scales the last term in \eqref{fullbeast} is
negligible compared to the first term. We obtain
$$ \frac{k_x^2}{k_z^2}\frac{N^2}{\Omega^2}\frac{s}{s_\kappa} \ll 1$$
For the fastest growing mode $s \lesssim q\Omega$ and $k_x/k_z \sim 1/q$,
and this condition reduces to
$$ k^2 \gg \frac{1}{q}\frac{N^2}{\Omega^2}
\left(\frac{\Omega}{\kappa}\right).$$
For there to exist $k$ that both satisfy this constraint in addition
to Eq.~\eqref{short}
 we must have
\begin{equation}
\frac{\nu}{\kappa}=\text{Pr} < q^2\frac{\Omega^2}{N^2},
\end{equation}
which is not a very onerous constraint at all, certainly not in PP
disks. 

In summary, on sufficiently long lengthscales viscosity is negligible, and on
sufficiently short lengthscales buoyancy is negligible. Thus 
the VSI exhibits classical double diffusive behaviour on the range
$$ q^{-1/2}\frac{N}{\Omega}k_\text{th}\, \ll\, k\, \ll\, q^{1/2}
k_\text{visc}, $$
and the VSI growth rate is almost exactly the same as if there were no
viscosity, no buoyancy, nor thermal diffusion. Note in the above that 
$k_\text{th}= \sqrt{\Omega/\kappa}$ and equals the (long) thermal
diffusion length and $k_\text{visc}=\sqrt{\Omega/\nu}$ and equals the
(very short) viscous diffusion length. 

In the Newtonian cooling case, we may ask the question for what
cooling rates yield VSI growth comparable to the unstratified
case. Following a similar procedure to earlier, 
we find that the second term
in \eqref{fullbeast} is negligible if 
\begin{equation}
\Omega\tau \ll q\left(\frac{\Omega^2}{N^2}\right).
\end{equation}
(To reach this conclusion we must initially assume the weaker
condition $\Omega\tau \ll 1/q$.)
Growth occurs at the same rate until
the viscous cut-off, which can be computed by setting $s=0$ and
solving for $k$. A bi-quadratic ensues and the critical $k$ is
\begin{equation}
k^2_\text{crit} = \frac{1}{2}\left(-n^2\tau\Omega + \sqrt{n^4(\tau\Omega)^2+4q^2}\right)\frac{\Omega}{\nu},
\end{equation}
to leading order in small $q$. (In deriving the above, we have assumed
that $k_x/k_z=-1/q$.) This wavenumber is generally quite
large, unless the cooling time is long, in which case we have the approximation
\begin{equation}
k_\text{crit} \approx \frac{q}{n\sqrt{\tau\Omega}}\sqrt{\frac{\Omega}{\nu}}, 
\end{equation}
which may yield small values. Instability is fully quenched when this wavenumber
equals $1/H$, furnishing us with an equation for $\tau$. The critical
cooling time above which the VSI dies is $(q^2/n^2)\text{Re}$. Note
that in the inviscid limit ($\text{Re}\to \infty$) instability is always
present no matter what the value of $\tau$. However, for long cooling
times the growth rates are too small to be interesting.

\section{Asymptotic long-time evolution of non-axisymmetric VSI modes}

It is convenient to employ units for which $\Omega_0=1$ and $k_y=1$, and
to introduce new dependent variables: $u=\bar{u}_x+2q\bar{u}_z$, $v=\bar{u}_y$,
$w=\bar{u}_x-2q\bar{u}_z$. We examine the three components of
\eqref{nonaxe}
in the limit of $q\ll 1$ and for long times: $t\gg 1/q$. To
ease the asymptotic ordering we set $t=\mathcal{O}(1/q^2)$ for the moment. 
Recognising that $u\sim w$, Eq.~\eqref{nonaxe} becomes
\begin{align}\label{ap1}
\frac{du}{dt} &= -\frac{2}{3t}u - \frac{16q^2}{9}v - \frac{q c_1}{t}v, \\
\frac{dv}{dt} &= -\frac{4}{3t}v -\frac{1}{2}w, \\
\frac{dw}{dt} &= -\frac{2}{3t}u + \frac{32q^2}{9}v - \frac{q c_2}{t}v,\label{ap3}
\end{align}
to leading order,
where $c_1=(8/9)k_z^0 - (16/27)q k_x^0$ and $c_2=-(40/9)k_z^0 +
(80/27)q k_x^0$ are constants depending on the initial wavevector. 
These equations may be reduced to a single ODE for $v$,
which to leading order is
\begin{equation} \label{3rd}
\frac{d^3v}{dt^3} + \frac{3}{t}\frac{d^2v}{dt^2}
+\left(\frac{16q^2}{9}- \frac{q c_2}{2t}\right)\frac{dv}{dt}
+ \frac{32q^2}{9t}v =0.
\end{equation}
The problem exhibits three time-scales, the fast orbital time $\sim
1$, a moderately slow time associated with the transient growth
of the non-axisymmetric VSI $\sim 1/q$, and a very slow time,
associated with its decay $\sim 1/q^2$. As a mode can exhibit
behaviour
on these differing time scales 
we adopt a formal multiscale approach. 
We are mainly interested in the longest times, as this will establish
stability or not, and thus define the very slow
time variable $T= q^2t$, where $T$ is order 1. There exists also a separate
intermediate time variable $\tau \sim q t$. Next the solution is
expanded in small $q$ so that $v= v_0(\tau,\,T) + q v_1(\tau,\,T)+ \dots$. Time
derivatives of this solution may be re-expressed as $d/dt= q \d/\d\tau +
q^2 \d/\d T$.  

Putting these assumptions and definitions into \eqref{3rd} then
collecting the various orders
in $q$ gives 
\begin{equation}
\mathcal{D}(v_0)\equiv\frac{\d^3v_0}{\d\tau^3} + \frac{16}{9}\frac{\d v_0}{\d\tau} =0,
\end{equation}
at $\mathcal{O}(q^3)$. We may then write down $v_0$ as a linear combination
of $\mathcal{D}$'s eigenfunctions:
\begin{equation}
v_0= A(T)\cos(\tfrac{4}{3}\tau) + B(T)\sin(\tfrac{4}{3}\tau) + C(T).
\end{equation}
Here $A$, $B$, and $C$ are amplitude functions, to be determined.

At the next order $\sim q^4$ we get
\begin{align*}
&\mathcal{D}(v_1)= \frac{32}{9}\left[\frac{dA}{dT} + \frac{A}{2T}+
  \frac{3c_2B}{16T}\right]\cos(\tfrac{4}{3}\tau) \notag\\ 
& +\frac{32}{9}\left[\frac{dB}{dT} + \frac{B}{2T}-
  \frac{3c_2A}{16T}\right]\sin(\tfrac{4}{3}\tau)-\frac{16}{9}\left[\frac{dC}{dT}+\frac{2C}{T}\right].
\end{align*}
Solvability of this equation requires that the right-hand side is
orthogonal to the eigenfunctions of $\mathcal{D}$. This simply means
that the expressions in square brackets must be zero, yielding two
first order ODEs for $A$ and $B$, solvable in terms of
power laws $\propto T^{\alpha}$. It is easy to show that 
$\alpha = -(1/2) \pm (3c_2/16)\text{i}.$ 
Meanwhile $C\propto T^{-2}$ and may be neglected as it decays at a
faster rate than $A$ or $B$.
Thus on the long timescale, $v$ always decays algebraically.
One can then go on to prove that the other velocity components
$u$ and $w$ decay no slower than $v$.

\end{appendix}

\end{document}